\documentclass[10pt,a4paper]{article}
 \pdfoutput=1
\usepackage{amsmath,fourier,amssymb,amsthm,graphicx,epstopdf,
color,cases}
\usepackage{chngcntr}
\counterwithout{figure}{section}
\usepackage[T1]{fontenc}
\allowdisplaybreaks
\numberwithin{equation}{section}

 \theoremstyle{plain}
 \newtheorem {hypo}{\bf\hspace{-\parindent}Hypothesis}[section]

 \newtheorem {lemma}[hypo]{Lemma}
 \newtheorem {theo}[hypo]{Theorem}
 \newtheorem {defin}[hypo]{Definition}
 \newtheorem {cor}[hypo]{Corollary}

 \theoremstyle{remark}
 \newtheorem {rmk}[hypo]{Remark}
  
 \newcommand{\pf}{\begin{bpf}}

 \newcommand{\pfms}{\begin{bpfms}}
 \newcommand{\epf}{\end{bpf}\hfill$\square$\vspace{0.1cm}}
 \newcommand{\epfms}{\end{bpfms}\hfill$\square$\\ }
 \newcommand\ben{\begin{equation*}}
 \newcommand\ebn{\end{equation*}}
 \newcommand\beq{\begin{equation}}
 \newcommand\eeq{\end{equation}}
 \newcommand\ds{\displaystyle}
  \newcommand\lb{\left(}
  \newcommand\rb{\right)} 
  
   \newcommand\Cb{\mathbb{C}} 
   \newcommand\Zb{\mathbb{Z}}
   \newcommand\Pb{\mathbb{P}} 
   
   \newcommand\bt{\mathbf{t}}

\begin{document}
\LARGE
\noindent
\textbf{Tau functions 
as Widom constants}
\normalsize
 \vspace{1cm}\\
 \noindent\textit{M. Cafasso$\,^{a,}$\footnote{cafasso@math.univ-angers.fr},
 P. Gavrylenko$\,^{b,c,d,}$\footnote{pasha145@gmail.com}, 
 O. Lisovyy$\,^{e,}$\footnote{lisovyi@lmpt.univ-tours.fr}}
 \vspace{0.2cm}\\
    $^a$ LAREMA, Université d'Angers, 2 bd Lavoisier, 49045 Angers, France
   \vspace{0.1cm}\\
 $^b$ Center for Advanced Studies, Skolkovo Institute of Science and Technology, 143026 Moscow, Russia
  \vspace{0.1cm}\\
 $^c$ National Research University Higher School of Economics, Department of
 Mathematics and International Laboratory of Representation Theory and Mathematical
 Physics, 119048 Moscow, Russia\vspace{0.1cm}\\
 $^d$ Bogolyubov Institute for Theoretical Physics,  03680 Kyiv, Ukraine
 \vspace{0.1cm}\\
 $^e$ Laboratoire de Math\'ematiques et Physique Th\'eorique CNRS/UMR 7350,  Universit\'e de Tours, Parc de Grandmont,
  37200 Tours, France

 \begin{abstract}
 We define a tau function for a generic Riemann-Hilbert problem  posed on a union of non-intersecting smooth closed curves with jump matrices analytic in their neighborhood. The tau function depends on parameters of the jumps and is expressed as the Fredholm determinant of an integral operator with block integrable kernel constructed in terms of elementary parametrices. Its logarithmic derivatives  with respect to parameters are given by contour integrals involving these parametrices and the solution of the Riemann-Hilbert problem. In the case of one circle, the tau function coincides with Widom's determinant arising in the asymptotics of block Toeplitz matrices. Our construction gives the Jimbo-Miwa-Ueno tau function for Riemann-Hilbert problems of isomonodromic origin (Painlev\'e VI, V, III, Garnier system, etc) and the Sato-Segal-Wilson tau function for integrable hierarchies such as Gelfand-Dickey and Drinfeld-Sokolov. 
 \end{abstract}

    \section{Introduction}
  Tau functions play a central role in the theory of integrable equations, both in fields of isospectral and isomonodromic deformations. They had been introduced in the 80s by the Kyoto school, with the explicitly stated aim \cite{JMU} to construct a generalization of the theta functions appearing, since Riemann \cite{Riemann}, as particular solutions of some non--linear equations\footnote{See also the work of S. Kovalevskaya \cite{Kov} and the more recent ones on finite-gap integration; e.g. \cite{DMN,Matv} and references therein.}.
  
  In the theory of isomonodromic deformations, tau functions are constructed starting from a certain differential 1-form $\omega_{\text{JMU}}$ defined on the space of the deformation parameters \cite{JMU}. Under the hypothesis that the parameters are of isomonodromic type, the form $\omega_{\text{JMU}}$ is closed and the tau function $\tau_{\text{JMU}}$ is defined (locally and up to a multiplicative constant) by the formula 
  \beq
  	\mathrm{d}\ln\tau_{\text{JMU}} := \omega_{\text{JMU}},
  \eeq 
  where $\mathrm{d}$ denotes the total differential. 
  
  Quite differently, on the side of isospectral deformations, Sato \cite{Sato} defined the tau function starting from his interpretation of the KP hierarchy in terms of the geometry of Grassmannian manifolds. Namely, to each solution of the KP hierarchy, one can associate a point $W$ in an infinite dimensional Grassmannian, and the related tau function is nothing but the formal series
  \beq
  	\tau_W := \sum_{\mathsf{Y} \in \mathbb{Y}} s_{\mathsf{Y}} W_{\mathsf{Y}}, 
  \eeq 
    where $ \mathbb{Y}$ is the set of partitions, $\big\{s_{\mathsf{Y}\in \mathbb{Y}}\big\}$ are the Schur polynomials and $\big\{ W_{\mathsf{Y}\in \mathbb{Y}}\big\}$ is the set of the Pl\"ucker coordinates of $W$. In \cite{SW}, Segal and Wilson provided an analytic version of Sato's theory, where formal series are replaced by $L^2$ functions, and rewrote the tau function as the (analytically well-defined) Fredholm determinant of a certain projection operator.
    
  Since the 80s, many generalizations of both definitions had been constructed; giving a complete account of the literature on the subject is out of the scope of this introduction. The generalizations touch different branches of mathematics as diverse as the representation theory of infinite--dimensional Lie algebras \cite{Kac}, Frobenius manifolds \cite{DZ}, instanton counting \cite{Nekrasov,NO}, Riemann--Hilbert boundary value problems \cite{Bertola,ILP} and topological recursion \cite{Eyn}, to name few of them. The reasons of such a flourishing literature, from our point of view, are to be found on the side of applications: while the several different definitions of tau functions could seem very abstract, the explicit computation of some of them are important for a growing mathematical community working on e.g. random matrix theory, statistical models, algebraic and symplectic enumerative geometry.
  
  The aim of this paper is to show that, at least for a very wide array of examples touching both the worlds of isomonodromic and isospectral deformations, tau functions coincide with a pretty simple object whose introduction by Widom goes back to 1976 \cite{Widom2}, before the very first seminal papers of the Kyoto school. Namely, they are the (Szeg\H o-)Widom constants associated to matrix--valued symbols  $J:\mathcal C\to\mathrm{GL}\lb N,\Cb\rb$, where $\mathcal C\subset\Cb\Pb^1$ is a circle centered at the origin. Recall that, given a symbol $J\lb z\rb=\sum_{k\in\Zb}J_k z^k$, the associated $n$-th block Toeplitz matrix is defined by
  $T_n\left[J\right]:=\lb J_{k-l}\rb_{k,l=1}^n$. 
       The asymptotics of $T_n[J]$ had been extensively studied in the operator theory literature, very often with motivations coming from applications in statistical mechanics such as, for instance, the Ising model \cite{DIK}.
       In particular, a celebrated theorem of Widom \cite{Widom2} states that, under certain analytical conditions on the symbol $J$, 
    \ben
    \lim_{n\to\infty}{G\left[J\right]}^{-n}\operatorname{det}T_n\left[J\right]=\tau\left[J\right].
    \ebn
    Here
    $G\left[J\right]=\exp \ds\frac{1}{2\pi i}\oint_{\mathcal C}
    z^{-1}\ln\operatorname{det}J\lb z\rb dz$
    and $\tau\left[J\right]$, which is known nowadays as the Widom constant, is the Fredholm determinant (the notations are explained in the next section)
    \beq\label{introtauRHP}
    \tau\left[J\right]= \operatorname{det}_{H_+}\lb \Pi_+ J^{-1}\Pi_+
    J\rb.
    \eeq 
    Indeed, this is a highly non-trivial extension of the celebrated strong Szeg\H o theorem \cite{Sz}, treating the case of scalar Toeplitz determinants.
    
    On the isospectral side, the coincidence between the Widom constant and the Sato-Segal-Wilson tau function had been established, for the so-called Gelfand-Dickey hierarchies, by one of the authors in \cite{Cafasso}, and successively extended to the Drinfeld-Sokolov hierarchies associated to an arbitrary Kac-Moody algebra in \cite{CafassoWu1}. Based on this identification, effective computations had been carried out in \cite{CafassoWu2,CDD} for topological and polynomial tau functions.
    
    In this work we show that, quite surprisingly, the recent results of \cite{GL16,GL17}, inspired by the isomonodromy/CFT/gauge theory correspondence, lead to a Fredholm determinant representation of the isomonodromic tau functions which is ultimately the same as given in \eqref{introtauRHP}. This implies, in particular, that the combinatorial expansion of the Sato's tau function and the much more recent series representations for the isomonodromic tau functions of Painlevé VI, V and III equations \cite{GIL12,GIL13} (at $0$), which originated from the AGT correspondence, are both of the same nature. Namely, they are nothing but the expansions of the Fredholm determinant \eqref{introtauRHP}, their terms being  products of Pl\"ucker coordinates of subspaces in the Sato-Segal-Wilson Grassmannian. 
    
    It turns out to be very fruitful to consider the symbol $J\lb z\rb$ as a jump matrix for a pair of Riemann-Hilbert problems (RHPs) on the circle $\mathcal C$. To construct Fredholm determinant and series representations for the tau functions of more general isomonodromic problems (e.g. the Garnier system), one needs to consider RHPs set on a union of non-intersecting ovals. In the present paper, we show how the definition \eqref{introtauRHP} of $\tau\left[J\right]$ can be generalized in this case and prove a formula for the log-differential of the appropriate extension of the Widom's determinant with respect to parameters of the jumps, which leads to its identification with the Jimbo-Miwa-Ueno tau function for RHPs of isomonodromic origin.

  We expect that the identification between the Widom constants and isomonodromic tau functions will lead to an effective way to compute the latter in the so far unsolved problems. These include, in particular, the construction of explicit asymptotic expansions of irregular type (at $\infty$) for Painlevé~I--V transcendents, cf \cite{BLMST,Nagoya}. Furthermore, the results of \cite{BSh,JNS} on $q$-Painlevé III and $q$-Painlevé VI equation give a hope that our approach may also be adapted to the $q$-difference setting.
    
     Starting from the foundational work \cite{Jimbo}, the standard scheme of asymptotic analysis of Painlevé transcendents \cite{FIKN} is to construct an \textit{approximate} solution of the appropriate RHP from solutions of ``elementary'' RHPs (parametrices), and then to extract the asymptotics of Painlevé functions from this approximation. The main ideological shift of our approach is that it gives an \textit{exact} Fredholm determinant expression for the tau functions in terms of parametrices, which define the relevant integrable kernels. The Fredholm determinant yields, with a relatively little effort, a complete asymptotic expansion of the tau function. The solution of the RHP (exact or approximate) is not needed at all, even though it can also be expressed via the resolvent of the appropriate integral operator. 
  
   Let us now briefly describe the organization of the paper. After introducing basic notations and recalling relevant results in Subsection \ref{subsecwidom}, we show that the Widom's determinant $\tau\left[J\right]$ admits a combinatorial series expansion whose individual terms are indexed by tuples of Young diagrams and are given by products of minors of Cauchy-Plemelj operators. In Subsection~\ref{subsec_appl}, we explain how $\tau\left[J\right]$ appears in the isomonodromy theory considering the example of Fuchsian systems with 4 regular singular points and systems with 2 irregular singularities; relations to previously known results on integrable hierarchies are also discussed. Section \ref{secmulti} is devoted to Riemann-Hilbert problems posed on a union of non-intersecting smooth closed curves. Specifically, we propose an extension of $\tau\left[J\right]$ to this case (Definition~\ref{defmultitau}) and establish the differentiation formulae for the so defined tau function with respect to parameters (Theorem~\ref{theomultitau}).

  \section{One-circle case\label{seconecircle}}
  \subsection{Widom formulas\label{subsecwidom}}
  Let $\mathcal C\subset\Cb\Pb^1$ be an anticlockwise oriented cicle centered at the origin, and let $f^{[+]}$ and $f^{[-]}$ denote its interior and exterior. Pick a loop $J:\mathcal C\to\mathrm{GL}\lb N,\Cb\rb$ that can be analytically continued into a fixed annulus $\mathcal A\supset \mathcal C$ and such that $\operatorname{det}J\lb z\rb$ has no winding along $\mathcal C$.
  We are going to associate to the pair $\lb \mathcal C,J\rb$ two  Riemann-Hilbert problems (RHPs). They ask to find $ \mathrm{GL}\lb N,\Cb\rb$ matrix functions $\Psi_{{\pm}}\lb z\rb$, $\bar\Psi_{{\pm}}\lb z\rb$ analytic in $f^{[\pm]}$ whose boundary values on $\mathcal C$ satisfy
  \begin{subequations}\label{RHPs}
  \begin{align}
  \label{RHPdirect}
  \text{direct RHP}:&\quad J\lb z\rb={\Psi_-\lb z\rb}^{-1} \Psi_+\lb z\rb,\\
  \label{RHPdual}
  \text{dual RHP}:&\quad  J\lb z\rb= \bar\Psi_+\lb z\rb{\bar\Psi_-\lb z\rb}^{-1}.
    \end{align}
  \end{subequations}
  It is a classical fact that $J\lb z\rb$ admits Birkhoff factorizations
  \beq\label{birkhoff}
  J\lb z\rb={Y_-\lb z\rb}^{-1} z^D Y_+\lb z\rb={\bar Y_+\lb z\rb} z^{\bar D} {\bar Y_-\lb z\rb}^{-1},
  \eeq
  where $Y_{\pm}\lb z\rb $, $\bar Y_{\pm}\lb z\rb $ can be continued to analytic functions in
  $f^{[\pm]}\cup \mathcal A$, and $D=\operatorname{diag}\lb d_1,\ldots,d_N\rb$, $\bar D=\operatorname{diag}\lb\bar d_1,\ldots,\bar d_N\rb$  with all $d_k,\bar d_{k'}\in\Zb$ such that $\sum_{k=1}^Nd_k=\sum_{k=1}^N\bar d_k=0$. The sets $\{d_k\}$ and $\{\bar d_k\}$ of partial indices  are uniquely determined by $J$.  The direct (dual) RHP is solvable iff $D=0$ (resp. $\bar D=0$).
  
  Introduce the Hilbert space $H=L^2\lb \mathcal C,\Cb^N\rb$. Its elements will be regarded as column vector functions. This space can be decomposed as $H=H_+\oplus H_-$, where the functions from $H_+$ ($H_-$)  continue analytically inside $\mathcal C$ (resp. outside $\mathcal C$ and vanish at $\infty$). We denote by $\Pi_{\pm}$ the projections on $H_{\pm}$ along $H_{\mp}$. 
  
  \begin{defin}\label{tauRHPdef} The tau function of the RHPs defined by $\lb \mathcal C,J\rb$ is defined as Fredholm determinant
  \beq\label{tauRHP}
  \tau\left[J\right]=\operatorname{det}_{H_+}\lb \Pi_+ J^{-1}\Pi_+
  J\rb.
  \eeq
  \end{defin}
  The operator $\Pi_+ J^{-1}\Pi_+
    J$ is known to be a trace class perturbation of the identity on $H_+$, which makes the determinant (\ref{tauRHP}) well-defined. The dual RHP is solvable iff the operator $P:=\Pi_+ J^{-1}$ is invertible on $H_+$, in which case its inverse is given by
  $P^{-1}=\bar\Psi_+\Pi_+ \bar\Psi_-^{-1}$. Likewise, the direct RHP is solvable iff the operator $Q:=\Pi_+ J$ is invertible, the inverse being equal to $Q^{-1}=\Psi_+^{-1}{\Pi_+ \Psi}_-$. If the direct or dual RHP is not solvable, then either $P$ or $Q$ has a nontrivial kernel and  $\tau\left[J\right]$ clearly vanishes. 
  
  Suppose that $J\lb z\rb$ admits a direct factorization (\ref{RHPdirect}). Define two Cauchy-Plemelj operators on $H$,
  \ben
  \mathsf a_H=\Psi_+\Pi_+\Psi_+^{-1}-\Pi_+,\qquad \mathsf d_H=\Psi_-\Pi_-\Psi_-^{-1}-\Pi_-.
  \ebn
  They can be explicitly written as integral operators 
  \ben
  \lb \mathsf a_H g\rb\lb z\rb=\frac{1}{2\pi i}\oint_{\mathcal C} \mathsf a\lb z,z'\rb g\lb z'\rb dz',\qquad 
  \lb \mathsf d_H g\rb\lb z\rb=\frac{1}{2\pi i}\oint_{\mathcal C} \mathsf d\lb z,z'\rb g\lb z'\rb dz',
  \ebn
  where
  \beq\label{adops}
  \mathsf a\lb z,z'\rb=\frac{\mathbb 1-\Psi_+\lb z\rb
  {\Psi_+\lb z'\rb}^{-1}}{z-z'},\qquad
    \mathsf d\lb z,z'\rb=\frac{\Psi_-\lb z\rb
    {\Psi_-\lb z'\rb}^{-1}-\mathbb 1}{z-z'}.
  \eeq
  The integral kernels $\mathsf a\lb z,z'\rb$ and $\mathsf d\lb z,z'\rb$ have integrable form, are not singular on the diagonal $z=z'$ and extend to analytic functions on $\mathcal A\times \mathcal A$. Since $\operatorname{im} \mathsf a_H\subseteq H_+\subseteq \operatorname{ker}\mathsf a_H$, $\operatorname{im} \mathsf d_H\subseteq H_-\subseteq \operatorname{ker}\mathsf d_H$, it is convenient to consider the restrictions
  \ben
  \mathsf a =\mathsf a_H\bigl|_{H_-}:  H_-\to H_+,\qquad
  \mathsf d =\mathsf d_H\bigl|_{H_+}:  H_+\to H_-.
  \ebn
  
  \begin{lemma} If the direct RHP (\ref{RHPdirect}) is solvable, then  $\tau[J]$ admits Fredholm determinant representation 
   \beq\label{FR1c}
   \tau[J]=\operatorname{det}_H\lb\mathbb 1+L\rb,\qquad 
   L=\lb \begin{array}{cc}
   0 & \mathsf a \\ \mathsf d & 0
   \end{array}\rb\in\mathrm{End}\lb  H_+\oplus H_-\rb,
   \eeq
   where integral operators $\mathsf a$ and $\mathsf d$ have block integrable kernels defined by (\ref{adops}).
  \end{lemma}
  \pf We have
   \ben
   \operatorname{det}_H\lb\mathbb 1+L\rb=\operatorname{det}_{H_+}
   \lb\mathbb 1-\mathsf a\mathsf d\rb
   =\operatorname{det}_{H_+}\lb\mathbb 1-
   \Psi_+\Pi_+\Psi_+^{-1}\Psi_-\Pi_-
    \Psi_-^{-1}\rb=
    \operatorname{det}_{H_+}\lb\mathbb 1-
    \Pi_+ J^{-1}\Pi_- J\rb,
   \ebn
   where the last equality is obtained by conjugating the action on $H_+$ by multiplication by $\Psi_+$.  Replacing $\Pi_-=\mathbb 1-\Pi_+$ in the last expression, we obtain the determinant (\ref{tauRHP}). \epf
  
  Of course, an analog of the representation (\ref{FR1c}) can also be written for the dual factorization (\ref{RHPdual}). However, from the point of view of applications it is convenient to consider  the direct factorization as given; the relevant matrix functions $\Psi_{\pm}$ \textit{define} the jump $J$. The Fredholm determinant (\ref{FR1c}) then yields an explicit representation for $\tau\left[J\right]$, whereas the dual factorization remains to be found.

  Let us now recall a formula for derivatives of $\tau\left[J\right]$ with respect to parameters of the jump matrix. It appeared as an intermediate result in the work of Widom \cite{Widom1} in 1974 and was rediscovered in \cite{IJK} more than 30 years later. As we expain below, this result is a precursor of the Jimbo-Miwa-Ueno definition of the isomonodromic tau function \cite{JMU}.
  \begin{theo}\label{WidomII} Consider a smooth family of $\,\mathrm{GL}\lb N,\Cb\rb$-loops  $(z,t)\mapsto J\lb z,t\rb$  which depend on an additional parameter $t$ and admit both factorizations (\ref{RHPs}). Then
  \beq\label{dertauRH}
  \partial_t \ln\tau\left[J\right]=\frac1{2\pi i}\oint_{\mathcal C}\operatorname{Tr}\left\{J^{-1}\partial_tJ\left[\partial_z\bar\Psi_-\, {\bar\Psi_-}^{-1}+\Psi_+^{-1}\,\partial_z\Psi_+
  \right]\right\}dz.
  \eeq
  \end{theo}
  \pf Using  previously defined operators $P=\Pi_+J^{-1}$, $Q=\Pi_+ J$ as well as their inverses $P^{-1}=\bar{\Psi}_+\Pi_+\bar\Psi_-^{-1}$, $Q^{-1}=\Psi_+^{-1}\Pi_+\Psi_-$, we may write
  \begin{gather}
  \nonumber\partial_t\ln\tau\left[J\right]=
  \partial_t\ln\operatorname{det}_{H_+}PQ=
  \operatorname{Tr}_{H_+}\lb \partial_t P\,P^{-1}+Q^{-1}\partial_t Q\rb=\\
  \label{widom20}
  =\operatorname{Tr}_{H_+}\lb-\Pi_+ J^{-1}\partial_t J\,J^{-1}\Pi_+
  \bar{\Psi}_+\Pi_+\bar\Psi_-^{-1} +
   \Psi_+^{-1}\Pi_+\Psi_-\Pi_+\partial_t J
  \rb.
  \end{gather}
  Let us simplify this expression. First observe that since 
  $\Pi_+\bar\Psi_+\Pi_+=\bar\Psi_+\Pi_+$ and $\Pi_+\Psi_-\Pi_-=0$, 
  the expression under trace (recall that this is an operator on $H_+$!) can be rewritten as
  \ben
  -\Pi_+ J^{-1}\partial_t J\,
    \bar{\Psi}_-\Pi_+\bar\Psi_-^{-1} +
     \Psi_+^{-1}\Pi_+\Psi_+J^{-1}\partial_t J=
     \Pi_+ J^{-1}\partial_t J\,\lb
         \bar{\Psi}_-\Pi_-\bar\Psi_-^{-1}-\Pi_-\rb+
  \lb\Psi_+^{-1}\Pi_+\Psi_+-\Pi_+\rb J^{-1}\partial_t J .      
  \ebn
  The operators ${\mathsf a}'_{H}=\Psi_+^{-1}\Pi_+\Psi_+-\Pi_+$, $\bar{\mathsf d}_H=\bar\Psi_-\Pi_-\bar\Psi_-^{-1}-\Pi_-$ on $H$ have the properties $\operatorname{im} {\mathsf a}_H'\subseteq H_+$, $H_-\subseteq \operatorname{ker} \bar{\mathsf d}_H$. This allows to extend the trace in \eqref{widom20} to the whole space $H$ and to rewrite it as
   \beq\label{widom21}
  \partial_t\ln\tau\left[J\right]=
  \operatorname{Tr}_{H}\left\{
          J^{-1}\partial_t J\lb{\mathsf a}_H'+\bar{\mathsf{d}}_H\rb\right\}.
  \eeq
  Since $ J^{-1}\partial_t J$ is a multiplication operator, to compute the last trace it suffices to know expressions for the kernels $\mathsf a'\lb z,z'\rb$, $\bar{\mathsf d}\lb z,z'\rb$ of the integral operators $\mathsf a_H'$, $\bar{\mathsf d}_H$ along the diagonal $z=z'$. From the respective counterparts of the formulas (\ref{adops}) it follows that
  \ben
  \mathsf a'\lb z,z\rb={\Psi_+\lb z\rb}^{-1}
  \partial_z\Psi_+\lb z\rb,\qquad 
  \bar{\mathsf d}\lb z,z\rb=\partial_z\bar\Psi_-\lb z\rb 
  {\bar\Psi_-\lb z\rb}^{-1}.
  \ebn
  In combination with (\ref{widom21}), this yields the statement of the theorem.
  \epf
  \begin{rmk}
  Theorem~\ref{WidomII} and its proof above clearly remain valid if we replace the circle $\mathcal C$ of \cite{Widom1,IJK} by any simple closed curve and consider the loops $J$ that continue to its tubular neighborhood. Notice that the right side of (\ref{dertauRH}) is closely related to the Malgrange-Bertola 1-form \cite{Malgrange,Bertola}
  \ben
  \omega_{\mathrm{MB}}\lb\delta\rb:=\frac1{2\pi i}\oint_{\mathcal C}\operatorname{Tr}\left\{J^{-1}\delta J\, \Psi_+^{-1}\,\partial_z\Psi_+\right\}dz,
  \ebn 
  where $\delta$ is an arbitrary vector field on the space of parameters of $J$.
  The curvature of the latter form does not necessarily vanish. Its analog defined by (\ref{dertauRH}) is on the other hand closed by construction, i.e. the contributions to curvature from the direct and dual factorization counterbalance each other. 
  An attempt to relate the tau function defined by $\omega_{\mathrm{MB}}$ (in those cases where $\omega_{\mathrm{MB}}$ is closed) to Fredholm determinants with integrable kernels was made in \cite{Bertola17}. It corresponds to decomposition of the initial RHP into a sequence of auxiliary RHPs with simpler matrix structure of the jumps, but does not seem to us to be directly related to our construction.
  \end{rmk}
  
  \subsection{Combinatorial expansion\label{subseccombi}}
  In this subsection we briefly outline some of the results on expansions of the Fredholm determinant $\tau\left[J\right]$. While the previous works \cite{GL16,GL17} focus on RHPs of isomonodromic origin, the combinatorial structure remains the same for generic jump $J$.

  \subsubsection{Maya and Young diagrams} 
\begin{figure}[h!]
              \centering
              \includegraphics[height=7cm]{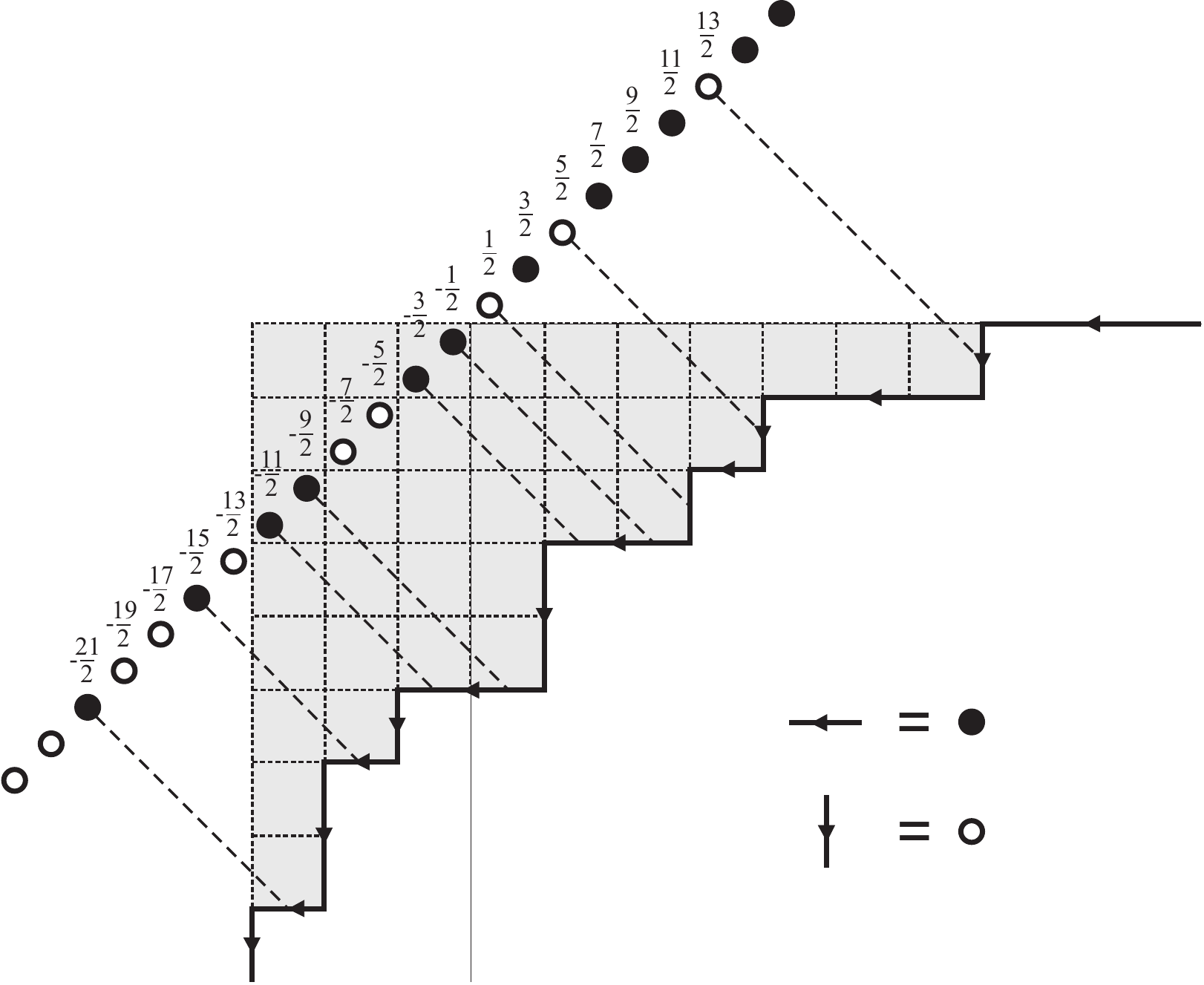}
              \begin{minipage}{0.77\textwidth}
              \caption{Correspondence between Maya and Young diagrams. The positions of particles and holes are $\mathsf p_X=\left\{\frac12,\frac52,\frac{13}2\right\}$ and
              $\mathsf h_X=\left\{-\frac{21}{2},-\frac{15}2,-\frac{11}2,
              -\frac92,-\frac32,-\frac12\right\}$. The charge $\mathsf Q_X=-3$ corresponds to signed distance between the vertical axis and left boundary of the profile of $\mathsf Y_X$.
                                 \label{Toeplitz_Fig1}}
                                                  \end{minipage}
                                \end{figure}

  Let $\Zb'=\Zb+\frac12$ be the half-integer lattice, $\Zb'_{\pm}=\Zb'_{\gtrless0}$, and let $\operatorname{Conf}\lb \Zb'\rb=\left\{0,1\right\}^{\Zb'}$ be the set of all finite subsets of $\Zb'$. The elements $X\subset\operatorname{Conf}\lb\Zb'\rb$  determine the positions of particles $\mathsf p_X:=X\cap\Zb'_{+}$ and holes $\mathsf h_X:=X\cap\Zb'_{-}$,
  thereby defining point configurations on $\Zb'$. A configuration $X$ may be alternatively represented by
  \begin{itemize} 
  \item A Maya diagram $\mathsf m_X$ obtained by drawing filled circles at 
  sites ${\lb\left.\Zb'_{+}\right\backslash \mathsf p_X\rb \cup \mathsf h_X}$ and empty circles
  at $\mathsf p_X\cup\lb\left.\Zb'_{-}\right\backslash \mathsf h_X\rb$, see Fig.~\ref{Toeplitz_Fig1}. The charge of $\mathsf m_X$ is defined as $\mathsf Q_X=\left| \mathsf p_X\right|-\left| \mathsf h_X\right|$. The set of all Maya diagrams will be denoted by $\mathbb M$.  
  \item A charged partition $\lb \mathsf Y_X,\mathsf Q_X\rb\in\mathbb Y\times \Zb$ where $\mathbb Y$ denotes the set of partitions $\mathsf Y=\lb\mathsf Y_1\ge\mathsf Y_2\ge\ldots \ge 0\rb$, with all ${\mathsf Y_k\in\Zb_{\ge0}}$. The partitions are identified with Young diagrams in the usual way. The Maya diagram corresponding to a charged partition $\lb \mathsf Y_X,\mathsf Q_X\rb$ can be described by the positions of empty circles, given by $\left\{\mathsf Y_k-k+\frac12+\mathsf Q_X\right\}_{k=1}^{\infty}$, cf Fig.~\ref{Toeplitz_Fig1}.
  \end{itemize}
  
  Let $L\in\Cb^{\mathfrak X\times\mathfrak X}$ be a matrix indexed by a discrete set $\mathfrak X$. The latter can be infinite, in which case $L$ is required to be a trace class operator on $\ell^2\lb\mathfrak X\rb$. The determinant $\operatorname{det}\lb \mathbb 1+L\rb$ can be expressed as the sum of principal minors enumerated by all possible subsets of $\mathfrak X$:
  \beq\label{vKf}
  \operatorname{det}\lb \mathbb 1+L\rb=\sum_{\mathfrak Y\in\left\{0,1\right\}^{\mathfrak X}}\operatorname{det} L_{\mathfrak Y},
  \eeq
  i.e. $L_{\mathfrak Y}$ is the restriction of $L$ to rows and columns $\mathfrak Y$. 
  
\begin{figure}[h!]
         \centering
         \includegraphics[height=5.5cm]{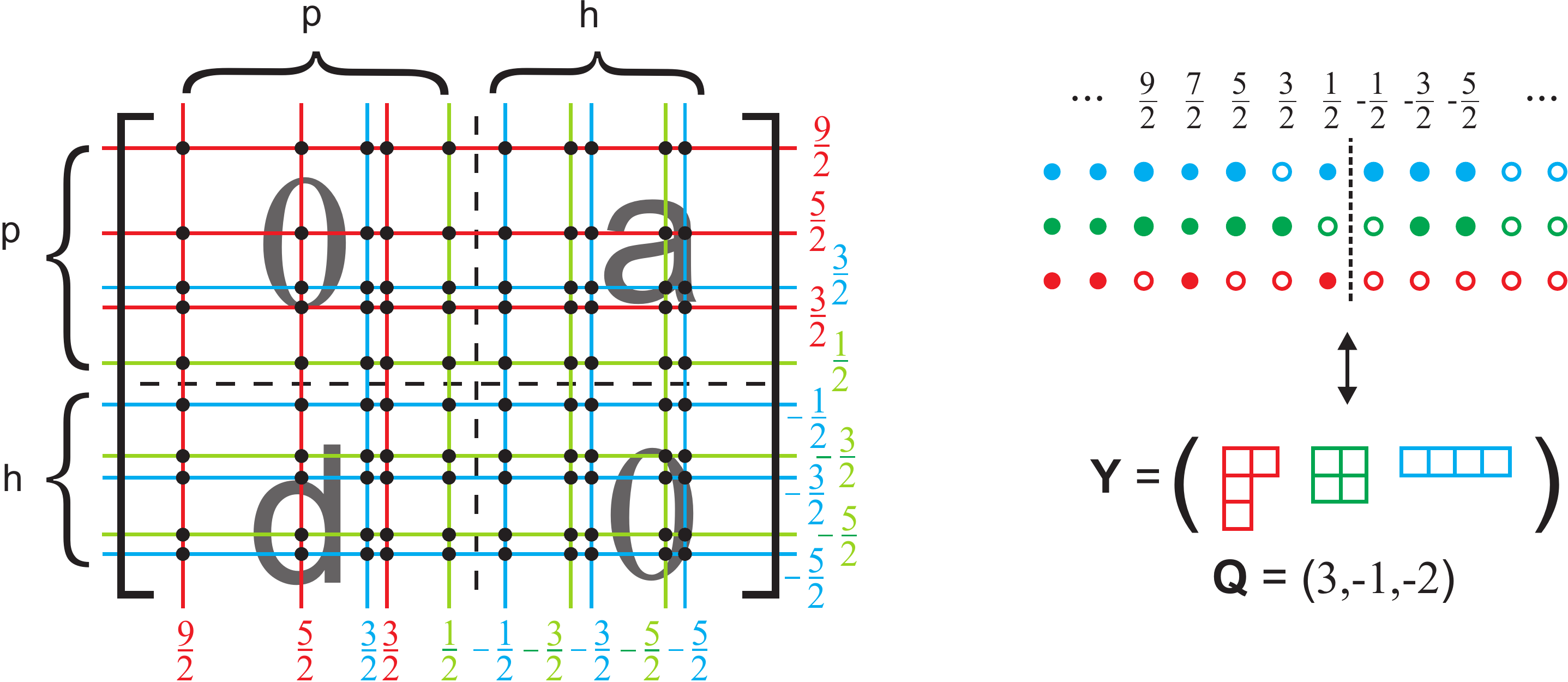}
         \begin{minipage}{0.77\textwidth}
         \caption{Example of labeling of principal minors with $N=3$ colors and $|\boldsymbol{\mathsf p}|=
         |\boldsymbol{\mathsf h}|=5$.
                            \label{Toeplitz_Fig2}}
                                             \end{minipage}
                           \end{figure}

  In order to apply this formula to the determinant (\ref{FR1c}), rewrite the integral operators $\mathsf a$ and $\mathsf d$ in the Fourier basis. Their kernels (\ref{adops}) may be expressed as
  \beq\label{fourierad}
  \mathsf{a}\lb z,z'\rb=\sum_{p,q\in\Zb'_{+}}\mathsf{a}^{\;\;\;\,p}_{-q}
  z^{-\frac12+p}z'^{-\frac12+q},\qquad
  \mathsf{d}\lb z,z'\rb=\sum_{p,q\in\Zb'_{+}}\mathsf{d}_{\;\;\;\,p}^{-q}
  z^{-\frac12-q}z'^{-\frac12-p},  
  \eeq
  where the coefficients $\mathsf{a}^{\;\;\;\,p}_{-q}, \mathsf{d}_{\;\;\;\,p}^{-q}$ are themselves $N\times N$ matrices whose elements we write as 
  $\mathsf{a}^{\;\;\;\,p;\alpha}_{-q;\beta}, 
  \mathsf{d}_{\;\;\;\,p;\beta}^{-q;\alpha}$. The ``color'' indices $\alpha,\beta=1,\ldots,N$ correspond to $\mathrm{GL}\lb N,\mathbb C\rb$-matrix structure of the RHP defined by the loop $J$. The principal minors of $L$ in (\ref{FR1c}) are therefore labeled by $N$-tuples of Maya diagrams 
  \ben
  \begin{gathered}
  \boldsymbol{\mathsf m}=\lb \mathsf m_1,\ldots,\mathsf m_N\rb=
  \bigl(\boldsymbol{\mathsf p},\boldsymbol{\mathsf h}\bigr)\in \mathbb M^N,\\
  \boldsymbol{\mathsf p}=\mathsf p_1\sqcup\ldots\sqcup \mathsf p_N,\qquad
  \boldsymbol{\mathsf h}=\mathsf h_1\sqcup\ldots\sqcup \mathsf h_N.
  \end{gathered}
  \ebn
  Here $\mathsf p_{\alpha}\in \left\{0,1\right\}^{\Zb'_{+}}$, $\mathsf h_{\alpha}\in \left\{0,1\right\}^{\Zb'_{-}}$ denote the positions of particles and holes of color $\alpha\in\left\{1,\ldots,N\right\}$.  The minors with $| \boldsymbol{\mathsf p}|\ne| \boldsymbol{\mathsf h}|$ clearly vanish, cf Fig.~\ref{Toeplitz_Fig2}. We may thus restrict the summation to $N$-tuples  of Maya diagrams  of zero total charge,
  \beq\label{combiexp}
  \begin{gathered}
  \tau\left[J\right]=\sum_{\boldsymbol{\mathsf m}\in\mathbb M^N:\,|\boldsymbol{\mathsf p}|=| \boldsymbol{\mathsf h}|}
 Z_{\;\boldsymbol{\mathsf m}}^{[+]}
  Z_{\;\boldsymbol{\mathsf m}}^{[-]},\\
   Z_{\;\boldsymbol{\mathsf m}}^{[+]}=\operatorname{det}
   \mathsf a^{\,\boldsymbol{\mathsf p}}_{\,\boldsymbol{\mathsf h}},\qquad
   Z_{\;\boldsymbol{\mathsf m}}^{[-]}=\lb -1\rb^{|\boldsymbol{\mathsf p}|} \operatorname{det}
     \mathsf d^{\,\boldsymbol{\mathsf h}}_{\,\boldsymbol{\mathsf p}}. 
  \end{gathered}
  \eeq

  The matrices $\mathsf a^{\,\boldsymbol{\mathsf p}}_{\,\boldsymbol{\mathsf h}},\mathsf d^{\,\boldsymbol{\mathsf h}}_{\,\boldsymbol{\mathsf p}}\in\operatorname{Mat}_{|\mathsf p|\times |\mathsf p| }\lb\Cb\rb$ correspond to the upper-right and lower-left block in the principal minor in Fig.~\ref{Toeplitz_Fig2}.
  Using the identification of Maya diagrams and charged partitions described above, the individual contributions to (\ref{combiexp}) may also be labeled by an $N$-tuple of partitions
  $\boldsymbol{\mathsf Y}\in\mathbb Y^N$ and an integer charge vector
  $\boldsymbol{\mathsf Q}\in \mathfrak Q_{N-1}$ from the $A_{N-1}$ root lattice
  \ben
  \mathfrak Q_{N-1}=\left\{\lb Q_1,\ldots,Q_N\rb\in\Zb^N\Bigl|\Bigr.\,\sum\nolimits_{\alpha=1}^NQ_{\alpha}=0\right\}.
  \ebn
  Adapting the notation, the combinatorial expansion (\ref{combiexp}) may then be written as
  \beq\label{combiexp2}
  \tau\left[J\right]=\sum_{\boldsymbol{\mathsf Q}\in\mathfrak Q_{N-1}}
  \sum_{\boldsymbol{\mathsf Y}\in\mathbb Y^N}
   Z_{\;\boldsymbol{\mathsf Y},\boldsymbol{\mathsf Q}}^{[+]}
    Z_{\;\boldsymbol{\mathsf Y},\boldsymbol{\mathsf Q}}^{[-]}.
  \eeq
  The structure of this series coincides with that of the dual Nekrasov-Okounkov partition functions introduced in \cite{NO}; in fact, in some cases these partition functions can be obtained as specializations of \eqref{combiexp2}.
   
  \begin{rmk}
  If $L$ is such that all principal minors $\operatorname{det}L_{\mathfrak Y}$ in (\ref{vKf}) are non-negative, then $\operatorname{Prob}\lb\mathfrak Y\rb:=\ds\frac{\operatorname{det}L_{\mathfrak Y}}{\operatorname{det}\lb \mathbb 1+L\rb}$ may be interpreted as a probability measure for a random point process on $\mathfrak X$ called the $L$-ensemble. This process is well-known to be determinantal and to have the correlation kernel $K=\ds\frac{L}{\mathbb 1+L}$. In our case, $L$ comes from a rewrite of the definition (\ref{tauRHP}) of $\tau\left[J\right]$ and
  \ben
  \mathfrak X\cong \left\{\boldsymbol{\mathsf m}\in\mathbb M^N:
  \,|\boldsymbol{\mathsf p}|=| \boldsymbol{\mathsf h}| \right\}\cong \mathfrak Q_{N-1}\times\mathbb Y^N.
  \ebn
   Explicit formulae for the inverses of $P$ and $Q$, already used in the proof  of Theorem~\ref{WidomII}, then allow to express $K \in\mathrm{End}\lb  H_+\oplus H_-\rb$ in terms of solutions of the direct and dual RHPs,
  \ben
  K=\lb\begin{array}{rr}
  \Pi_+\Psi_+\bar{\Psi}_-\Pi_- \bar{\Psi}_-^{-1}\Psi_+^{-1}\Pi_+ &
 - \Pi_+\Psi_+\bar{\Psi}_-\Pi_- \bar{\Psi}_-^{-1}\Psi_+^{-1}\Pi_- \\
 -\Pi_-\Psi_+\bar{\Psi}_-\Pi_+ \bar{\Psi}_-^{-1}\Psi_+^{-1}\Pi_+ &
 \Pi_-\Psi_+\bar{\Psi}_-\Pi_+ \bar{\Psi}_-^{-1}\Psi_+^{-1}\Pi_-
  \end{array}\rb.
  \ebn
  \end{rmk}
 Denote by  $\chi_{\mathfrak J} $ the indicator
 function of any subset $\mathfrak J \subseteq \mathbb{Z}'$.
 It is worth observing that the component $K_{++} := \chi_{\Zb_+'^N} K\chi_{\Zb_+'^N}$, i.e. the term in the upper-left corner of the matrix representation above, is nothing but the Fredholm operator appearing in the celebrated Borodin-Okounkov formula \cite{BO}, in its matrix version\footnote{For comparison, one should use that  $K_{++} = \Pi_+\Psi_-\bar{\Psi}_+\Pi_- \bar{\Psi}_-^{-1}\Psi_+^{-1}\Pi_+$ and compare the factorizations of the jump $\phi$ in \cite{BW} with those of $J^{-1}$ in the present paper. } \cite{BW}. Hence, the (gap) probability of finding no particles of any colour in the sites $\mathfrak J_n=\left\{k+\tfrac12\in\Zb',k\ge n\right\} $ is given by 
    \beq
    	\operatorname{Prob}\lb\mathbf{p} \mathop{\cap} \mathfrak J_n^N = \emptyset \rb = \det \lb\mathbb{1} - \chi_{\mathfrak J_n^N}K_{++}\chi_{\mathfrak J_n^N} \rb =  \frac{\det T_n\left[ J^{-1}\right]}{\tau\left[J\right]},
     \eeq
   where the last equality (valid under assumption that $\operatorname{det}J\lb z\rb$ has geometric mean $1$) is precisely the content of the Borodin--Okounkov formula.  
  
  \subsubsection{Grassmannian interpretation}\label{Grassmannians}
        It is possible to give an interpretation of the formulas above in the setting of the Sato-Segal-Wilson theory of infinite-dimensional Grassmannians. We will start with the analytic theory \cite{SW} and then comment on the relation with Sato's formal definition of tau function \cite{Sato}. Consider the point $W := \Psi_- \cdot H_+$ in the Segal-Wilson Grassmannian $\mathrm{Gr}\lb H\rb$.
    The subspace $W$ is spanned by the columns of the (rectangular) matrix
    \ben
            G^{[-]} := \begin{pmatrix} \Pi_+ \Psi_- \Pi_+ \\ \Pi_- \Psi_- \Pi_+\end{pmatrix}.
    \ebn
        This is a \emph{frame} for the point $W$. More generally, a frame for $W$ will be a rectangular matrix $(w_+, w_-)^{T}$ whose columns span $W$, and the frame will be called \emph{admissible} if $w_+ - \mathbb{1}$ is of trace class on $H_+$. Of course, in general $G^{[-]}$ will not be admissible. Nevertheless, since $\Pi_+ \Psi_- \Pi_+$ is invertible, there is a canonical way to transform $G^{[-]}$ into an admissible frame by right multiplication:
  \ben
        G^{[-]} \mapsto G^{[-]}\lb\Pi_+ \Psi_- \Pi_+\rb^{-1} = \begin{pmatrix} \mathbb{1}\\ \Pi_- \Psi_- \Pi_+\Psi^{-1}_- \Pi_+ \end{pmatrix} = \begin{pmatrix} \mathbb{1}\\ -{\mathsf d}\end{pmatrix}
  \ebn
  In other words, $-\mathsf{d}$ is the map whose graph is equal to $W$. Now, we can act on $W$ with multiplication by $\Psi_+^{-1}$, and the Segal-Wilson tau function $\tau_W\lb \Psi_+\rb$ is defined by the formula
  \beq
      \tau_W\lb\Psi_+\rb.\Psi_+^{-1}\sigma\lb W\rb = \sigma\lb\Psi_+^{-1}W\rb,
  \eeq
  where $\sigma$ is the canonical global section of the determinant line bundle $\mathrm{Det}^*$ over $\mathrm{Gr}\lb H\rb$, and the action of $\Psi_+^{-1}$ on $\mathrm{Gr}\lb H\rb$ is extended to  $\mathrm{Det}^*$. The reader is referred to \cite{SW} for the details. What is important here is that, since the operator of multiplication by $\Psi_+^{-1}$ has block form of type
  \ben
          \Psi_+^{-1} = \begin{pmatrix}  \Pi_+ \Psi_+^{-1} \Pi_+ & \Pi_+ \Psi_+^{-1} \Pi_-\\ 0 &  \Pi_- \Psi_+^{-1} \Pi_- \end{pmatrix},
  \ebn
  the tau function is given by the Fredholm determinant, see  \cite[formula (3.5)]{SW},
  \beq\label{SWtau}
      \tau_{W}\lb\Psi_+\rb = \mathrm{det}_{H_+}\lb\mathbb{1} - \Pi_+ \Psi_+ \Pi_+ \Psi_+^{-1} \Pi_-\mathsf{d}\rb = \mathrm{det}_{H_+}\lb\mathbb{1} - \mathsf{a}\mathsf{d}\rb,
  \eeq
  so that finally we have $\tau_W\lb \Psi_+\rb = \tau\left[J\right]$.
  
  If we are willing to work with formal series instead of analytic functions, there is no reason to restrict to admissible frames and, in Sato's style \cite{Sato}, one can simply define the tau function as
  \beq\label{Satotau1}
      \tau_W\lb\Psi_+\rb = \mathrm{det}_{H_+}\Big(G^{[+]}G^{[-]}\Big),
  \eeq
  where $G^{[+]}$ and $G^{[-]}$ are given, respectively, by the matrices associated to  $\Pi_+ \Psi_+^{-1}$ and  $\Psi_- \Pi_+$:
      \ben
          \Pi_+ \Psi_+^{-1} = G^{[+]} := \begin{pmatrix} \Pi_+ \Psi_+^{-1} \Pi_+ & \Pi_+ \Psi_+^{-1} \Pi_-\end{pmatrix}, \quad \quad \Psi_-\Pi_+ = G^{[-]} := \begin{pmatrix} \Pi_+ \Psi_- \Pi_+ \\ \Pi_- \Psi_- \Pi_+ \end{pmatrix}.
      \ebn
  Since $ \Pi_+ \Psi_+^{-1} \Pi_+$ and $ \Pi_+ \Psi_- \Pi_+$ are respectively upper and lower triangular with the identity on the main diagonal, the two definitions \eqref{SWtau} and \eqref{Satotau1} are (formally!) the same. Nevertheless, $G^{[+]}G^{[-]} - \mathbb{1}$ is not a trace class operator, therefore the determinant in \eqref{Satotau1} is to be understood as the limit of the determinant whose size goes to infinity. Indeed, this is nothing but $\operatorname{det}T_{n\to\infty}\left[J^{-1}\right]$, and the equality between \eqref{SWtau} and \eqref{Satotau1} is simply a rephrasing of the Widom's theorem stated in the introduction. The way to compute the latter is through the Cauchy-Binet formula. Namely, for any $N$-tuple $\boldsymbol{\mathsf m}$ of Maya diagrams  of zero total charge, define the associated Pl\"ucker coordinates $G^{[\pm]}_{\boldsymbol{\mathsf m}}$ as the determinants of the square matrices obtained by choosing the columns/lines of the matrix $G^{[\pm]}$ in correspondence with the filled circles in   $\boldsymbol{\mathsf m}$. Then $\tau_W\lb\Psi_+\rb$ is given by
  \beq\label{SatoTau}
      \tau_W\lb\Psi_+\rb = \sum_{\boldsymbol{\mathsf m}\in\mathbb M^N:\,|\boldsymbol{\mathsf p}|=| \boldsymbol{\mathsf h}|} G^{[+]}_{\boldsymbol{\mathsf m}}G^{[-]}_{\boldsymbol{\mathsf m}}.
  \eeq
  It is natural to wonder what is the relation between the two expansions \eqref{SatoTau} and \eqref{combiexp}. The answer is that they are actually the same, since      $G^{[\pm]}_{\boldsymbol{\mathsf m}} = 
  Z^{[\pm]}_{\boldsymbol{\mathsf m}}$.
  This identity (cf Proposition 2.1 in \cite{EH}) is indeed the main step in the proof of the so-called Giambelli identity, relating Pl\"ucker coordinates associated to an arbitrary Young diagram to the hooked ones. We conclude this subsection by noticing that the integral formula \eqref{adops} for the so-called affine coordinates $\mathsf{a}$, $-\mathsf{d}$ already appeared, in the context of Gelfand-Dickey equations, in \cite{BD}.

  \subsubsection{Matrix elements}
 The reader might wonder when matrix elements $\mathsf{a}^{\;\;\;\,p}_{-q}, \mathsf{d}_{\;\;\;\,p}^{-q}$ become effectively computable. In applications to integrable hierarchies, the entries of $\mathsf{a}$ are universal (i.e. they do not depend on the solution but just on the hierarchy) and are given explicitly in terms of the elementary Schur polynomials, while $\mathsf{d}$ determines the point in the Grassmannian corresponding to the given solution, see Subsection \ref{subsecIH} below for more details. Another typical situation where such calculation is possible occurs in the context of monodromy preserving deformations.  Suppose that $\Psi_{\pm}\lb z\rb$ satisfy 
  \beq\label{DEcond}
  \partial_z\Psi_{\pm}\lb z\rb= \Psi_{\pm}\lb z\rb A_{\pm}\lb z\rb+z^{-1}\Lambda_{\pm}\lb z\rb\Psi_{\pm}\lb z\rb ,
  \eeq
  with $A_{\pm}\lb z\rb$ rational in $z$ and $\Lambda_{\pm}\lb z\rb$ polynomial in $z^{\pm1}$. The latter condition holds in a number of examples where $\Psi_{\pm}\lb z\rb$ are related to fundamental matrix solutions of linear systems. It should be seen as an analog of the conditions used in \cite{TW93} to derive nonlinear PDEs satisfied by Fredholm determinants of certain scalar integrable kernels.
  
  Introduce the operator $\mathcal L_0=z\partial_z+z'\partial_{z'}+1$. Since $\mathcal L_0\frac{1}{z-z'}=0$, we have 
  \beq\label{rhsME}
  \mathcal L_0 \mathsf a_{\pm}\lb z,z'\rb=\pm\Psi_{\pm}\lb z\rb \mathbf A_{\pm}\lb z,z'\rb{\Psi_{\pm}\lb z'\rb}^{-1}+
   \Lambda_{\pm}\lb z\rb \mathsf a_{\pm}\lb z,z'\rb-
  \mathsf a_{\pm}\lb z,z'\rb  \Lambda_{\pm}\lb z'\rb\pm
  \boldsymbol{\Lambda}_{\pm}\lb z,z'\rb,
  \eeq
  where $\mathsf a_{+}\lb z,z'\rb=\mathsf a\lb z,z'\rb$, $\mathsf a_{-}\lb z,z'\rb=\mathsf d\lb z,z'\rb$ and
  \ben
  \mathbf A_{\pm}\lb z,z'\rb=\frac{
    z'A_{\pm}\lb z'\rb-zA_{\pm}\lb z\rb}{z-z'},\qquad
    \boldsymbol{\Lambda}_{\pm}\lb z,z'\rb=\frac{
        \Lambda_{\pm}\lb z'\rb-\Lambda_{\pm}\lb z\rb}{z-z'}.
  \ebn
  There exist $M_{\pm}\in\Zb_{\ge0}$ such that $\mathbf A_{\pm}\lb z,z'\rb=\sum_{m=1}^{M_{\pm}}\varphi_{m,\pm}\lb z\rb\otimes
  \bar\varphi_{m,\pm}\lb z'\rb$, where $\varphi_{m,\pm}$ and
   $\bar\varphi_{m,\pm}$ are column and raw $N$-vectors. On the other hand, applying $\mathcal L_0$ directly to Fourier expansions (\ref{fourierad}), one has 
   \beq\label{lhsME}
   \begin{aligned}
     \mathcal L_0\mathsf{a}\lb z,z'\rb=&\;\;\;\;\,\sum\nolimits_{p,q\in\Zb'_{+}}\lb p+q\rb \mathsf{a}^{\;\;\;\,p}_{-q}
     z^{-\frac12+p}z'^{-\frac12+q},\\
       \mathcal L_0\mathsf{d}\lb z,z'\rb=&-\sum\nolimits_{p,q\in\Zb'_{+}}\lb p+q\rb \mathsf{d}_{\;\;\;\,p}^{-q}
       z^{-\frac12-q}z'^{-\frac12-p}.   
   \end{aligned}  
   \eeq
  Comparing this expression with (\ref{rhsME}), we obtain a system of linear equations that determine $\mathsf{a}^{\;\;\;\,p}_{-q}$, $\mathsf{d}_{\;\;\;\,p}^{-q}$ in terms of Fourier modes of $\Psi_{\pm}\lb z\rb\varphi_{m,\pm}\lb z\rb$,
  $\bar\varphi_{m,\pm}\lb z\rb {\Psi_{\pm}\lb z\rb}^{-1}$ and the coefficients of $\Lambda_{\pm}\lb z\rb\in \operatorname{Mat}_{N\times N}\lb\Cb\left[z^{\pm1}\right]\rb$, $\boldsymbol{\Lambda}_{\pm}\lb z,z'\rb\in\operatorname{Mat}_{N\times N}\lb\Cb\left[z^{\pm1},z'{}^{\pm1}\right]\rb$.
  
   The simplest nontrivial situation  corresponds to $M_+=M_-=1$ and  $\Lambda_{\pm}$ given by constant diagonal matrices. It occurs, in particular, for generic (non-logarithmic) solutions of Painlevé VI, V and III. In these cases, (\ref{rhsME}) and (\ref{lhsME}) imply that 
   \ben
   \begin{aligned}
   \sum\nolimits_{p,q\in\Zb'_{+}}\lb p+q-\mathrm{ad}_{\Lambda_+}\rb\mathsf{a}^{\;\;\;\,p}_{-q}
        z^{-\frac12+p}z'^{-\frac12+q}=&\Psi_+\lb z\rb\varphi_{+}\lb z\rb\otimes
          \bar\varphi_{+}\lb z'\rb \Psi_+\lb z'\rb^{-1},\\
   \sum\nolimits_{p,q\in\Zb'_{+}}\lb p+q+\mathrm{ad}_{\Lambda_-}\rb\mathsf{d}_{\;\;\;\,p}^{-q}
        z^{-\frac12-q}z'^{-\frac12-p}=&\Psi_-\lb z\rb\varphi_{-}\lb z\rb\otimes
          \bar\varphi_{-}\lb z'\rb \Psi_-\lb z'\rb^{-1}.          
   \end{aligned}
   \ebn The modes $\mathsf{a}^{\;\;\;\,p}_{-q}$, $\mathsf{d}_{\;\;\;\,p}^{-q}$ are therefore given by Cauchy matrices which in turn implies that the determinants
   $Z_{\;\boldsymbol{\mathsf Y},\boldsymbol{\mathsf Q}}^{[\pm]}$ have nice factorized expressions.

  \subsection{Applications}\label{subsec_appl}
  In this subsection, we demonstrate the relation between our Definition~\ref{tauRHPdef} and tau functions of certain classes of isomonodromic systems and integrable hierarchies. The general strategy is to reduce the associated linear problem to a RHP on a circle and make use of Widom's differentiation formula (Theorem~\ref{WidomII}).
  
  \subsubsection{Four regular singularities}\label{subsec4reg}
  Our basic example deals with a linear system with four Fuchsian singularities placed at $0,t,1,\infty$. It is given by
  \beq \label{fuchs4}
  \partial_z\Phi=\Phi A\lb z\rb,\qquad A\lb z\rb=\frac{A_0}{z}+\frac{A_t}{z-t}+
  \frac{A_1}{z-1},
  \eeq
  with $ A_{0,t,1}\in\operatorname{Mat}_{N\times N}\lb \Cb\rb$.
  Consider generic situation where $A_{0,t,1}$ and $A_{\infty}:=-A_0-A_t-A_1$ are diagonalizable. For $a=0,t,1,\infty$, fix the diagonalizations $A_{a}=G_{a}^{-1}\Theta_{a} G_{a}$ with diagonal $\Theta_{a}$. 
   Assume that the eigenvalues of $A_{a}$ are distinct mod $\mathbb Z$. Then there exist unique fundamental matrix solutions $\Phi^{(a)}\lb z\rb$ of (\ref{fuchs4}), holomorphic on the universal covering  of $\Cb\backslash\left\{0,t,1\right\}$ and such that 
   \ben
   \Phi^{(a)}\lb z\rb=\begin{cases} \lb a-z\rb^{\Theta_{a}}G^{(a)}\lb z\rb,\qquad &\text{for }a=0,t,1, \\  \lb -z\rb^{-\Theta_{\infty}}G^{(\infty)}\lb z\rb,\qquad &\text{for } a=\infty,
  \end{cases}
  \ebn 
  where $G^{(a)}\lb z\rb$ is holomorphic and invertible in a finite open disk around $z=a$ and satisfies the normalization condition ${G^{(a)}\lb a\rb=G_{a}}$. 
  
  Further assume for notational simplicity that $t\in\lb 0,1\rb$.  The canonical solutions $\Phi^{(0,\infty)}\lb z\rb$ analytically continue to single-valued matrix functions on the cut Riemann sphere $\Cb\Pb^1\backslash\mathbb R_{\ge0}$. Similarly, $\Phi^{(t)}\lb z\rb$
  and $\Phi^{(1)}\lb z\rb$ are naturally defined on $\Cb\Pb^1\backslash\left((-\infty,0]\cup[t,\infty)\right)$ and 
  $\Cb\Pb^1\backslash\left((-\infty,t]\cup[1,\infty)\right)$, respectively.
  Take an arbitrary fundamental solution $\Phi\lb z\rb$, defined on $\Cb\Pb^1\backslash\mathbb R_{\ge0}$. The connection matrices 
  $C_{a,\epsilon}=\Phi\lb z\rb {\Phi^{(a)}\lb z\rb}^{-1}$, with  $\epsilon = \operatorname{sgn}\Im z$, are independent of $z$. They satisfy the compatibility conditions
  \beq\label{compa}
  \begin{gathered}
  \begin{aligned}C_{0,+}=C_{0,-},&\qquad C_{\infty,+}=C_{\infty,-},\\
  M_0=C_{0,-}e^{2\pi i\Theta_0}C_{0,+}^{-1}=C_{t,-}C_{t,+}^{-1},&\qquad M_{\infty}^{-1}=C_{1,-}e^{2\pi i\Theta_1}C_{1,+}^{-1}=
  C_{\infty,-}e^{-2\pi i\Theta_{\infty}}C_{\infty,+}^{-1},
  \end{aligned}\\
  M_0M_t=\lb  M_1 M_{\infty}\rb^{-1}=C_{t,-}e^{2\pi i\Theta_t}C_{t,+}^{-1}=C_{1,-}C_{1,+}^{-1}.\qquad
  \end{gathered}
  \eeq
  where $M_{a}$ denotes anticlockwise monodromy matrix of $\Phi\lb z\rb$ around the Fuchsian singular point $a\in\left\{0,t,1,\infty\right\}$. The connection matrices $\{C_{a,\pm}\}$ and exponents $\left\{\Theta_{a}\right\}$ of local monodromy constitute the monodromy data for the 4-point Fuchsian system (\ref{fuchs4}). 
  
\begin{figure}[h!]
   \centering
   \includegraphics[height=6.5cm]{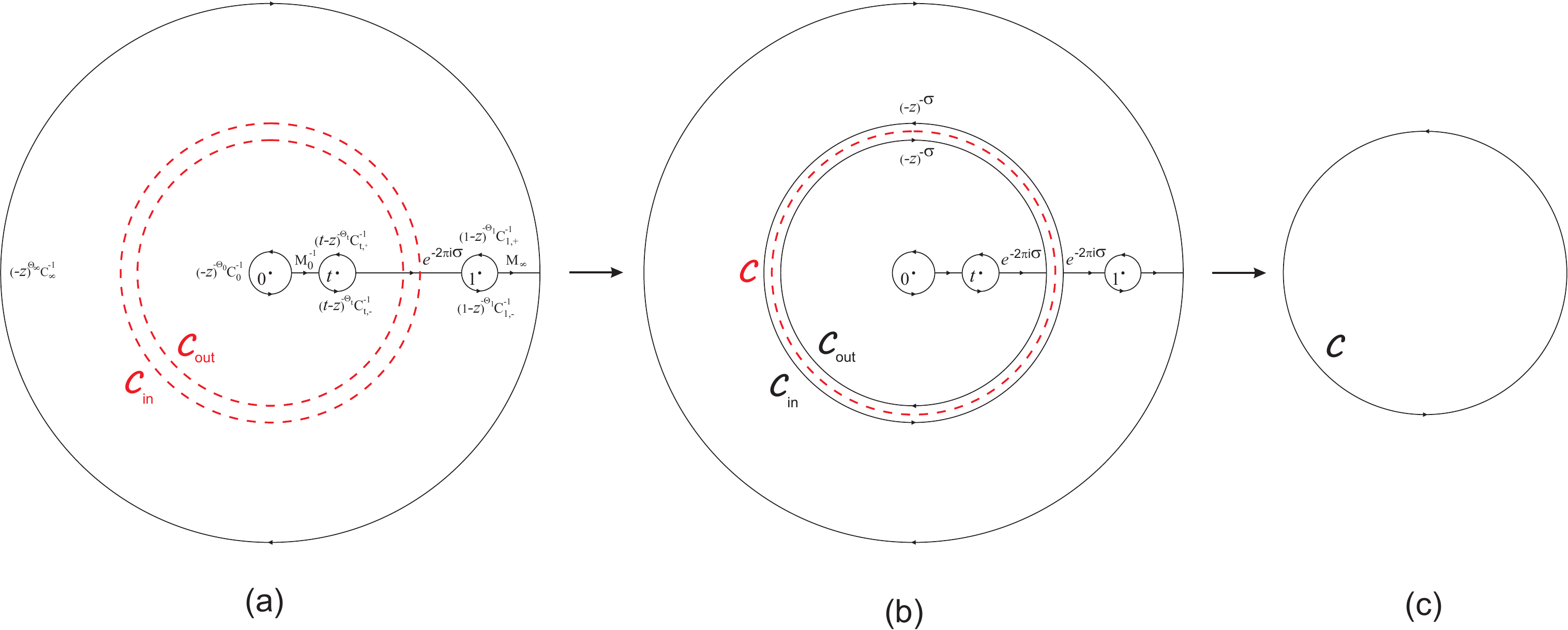}
   \begin{minipage}{0.77\textwidth}
   \caption{RH contours for (a) $\tilde \Psi$
   (b) $\hat \Psi$ (c) $\bar\Psi$.
                      \label{Toeplitz_Fig3}}
                                       \end{minipage}
                     \end{figure}

  Let us now explain how to transform (\ref{fuchs4}) into a Riemann-Hilbert problem on a circle. This will be achieved in several steps:
  \begin{enumerate}
  \item Start with the contour $\tilde{\Gamma}$ shown in Fig.~\ref{Toeplitz_Fig3}a by solid black curves. Denote by $D_{a}$ the disk around $z=a$ bounded by $\gamma_{a}$  and define
  \ben
  \tilde\Psi\lb z\rb=\begin{cases}
  G^{(a)}\lb z\rb,\qquad& z\in D_{a},\\
  \Phi\lb z\rb, \qquad &z\notin\mathbb R_{\ge0}\cup \bar D_0\cup \bar D_t\cup \bar D_1 \cup \bar D_{\infty}.
  \end{cases} 
  \ebn
  Comparing with (\ref{RHPdual}), we see that the matrix function $\tilde{\Psi}\lb z\rb$ solves a dual RHP set on $\tilde \Gamma$ with the jumps indicated in Fig.~\ref{Toeplitz_Fig3}a.
  \item Next cancel the constant jump $\lb M_0M_t\rb^{-1}$ on the real segment cut out by the dashed red circles $\mathcal C_{\mathrm{out,in}}$. To this end, let us write $M_0M_t=e^{2\pi i \mathfrak S}$. There is a certain freedom in the choice of  $\mathfrak S$;  for example, in the generic situation where  $\mathfrak S$ may be assumed diagonal, we may add to it any integer diagonal matrix.  Denote by $\hat{\mathcal A}$ the open annulus bounded by $\mathcal C_{\mathrm{out,in}}$ and set
  \ben
  \hat\Psi\lb z\rb=\begin{cases}
    \lb -z\rb^{-\mathfrak S}\tilde\Psi\lb z\rb,\qquad& z\in\hat{\mathcal A},\\
    \tilde\Psi\lb z\rb, \qquad &z\notin\bar{\hat{\mathcal A}}.
    \end{cases}
  \ebn
  The dual RHP for $\hat{\Psi}\lb z\rb$ is  set on the contour $\hat \Gamma$ indicated in Fig.~\ref{Toeplitz_Fig3}b by solid black lines. The jump matrices associated to $\mathcal C_{\mathrm{out}}$ and $\mathcal C_{\mathrm{in}}$ are $\lb -z\rb^{-\mathfrak S}$; on the rest of the contour the jumps are the same as for~$\tilde\Psi\lb z\rb$. 
  \item The contour $\hat\Gamma$ has two connected components, $\hat\Gamma_-$ and $\hat\Gamma_+$, containing respectively $\mathcal C_{\mathrm{out}}$ and $\mathcal C_{\mathrm{in}}$. Choose $\mathfrak S$ so that $\operatorname{Tr} \mathfrak S=\operatorname{Tr}\lb \Theta_0+\Theta_{t}\rb=-\operatorname{Tr}\lb \Theta_1+\Theta_{\infty}\rb$ (this choice still allows for  $\mathfrak Q_{N-1}$-shifts). The RHPs obtained by restricting the initial contour to $\hat \Gamma_-$ or $\hat\Gamma_+$ while keeping the same jumps  are then generically solvable. Their solutions are related to fundamental matrices $\Phi_-\lb z\rb$ and $\Phi_+\lb z\rb$ of 3-point Fuchsian systems whose singular poins are  $0,t,\infty$ and $0,1,\infty$. Let us denote these solutions by $\Psi_-\lb z\rb$ and $\Psi_+\lb z\rb$. The subscript reminds that these functions are analytic outside $\mathcal C_{\mathrm{out}}$ and inside $\mathcal C_{\mathrm{in}}$, respectively. 
  
  Consider an auxiliary circle $\mathcal C$ inside $\hat{\mathcal A}$, indicated by dashed red line in Fig.~\ref{Toeplitz_Fig3}b, and define
  \beq\label{psi1c}
  \bar\Psi\lb z\rb=
  \begin{cases}
  {\Psi_+\lb z\rb}^{-1}\hat\Psi\lb z\rb,\qquad & \text{outside }\mathcal C,\\
  {\Psi_-\lb z\rb}^{-1}\hat\Psi\lb z\rb,\qquad &
  \text{inside }\mathcal C.
  \end{cases}
  \eeq
  The matrix function $\bar\Psi\lb z\rb$ has no jumps except on  $\mathcal C$. The jump of the relevant \textit{dual} RHP is written in the form of \textit{direct} factorization,
  \beq\label{jump4fuchs}
  J\lb z\rb={\Psi_-\lb z\rb}^{-1}\Psi_+\lb z\rb,
  \eeq
  cf (\ref{RHPdirect}). The problem of solving the 4-point Fuchsian system with a prescribed monodromy is therefore converted into a RHP for $\bar\Psi_{\pm}\lb z\rb$ on a single circle (Fig.~\ref{Toeplitz_Fig3}c), with the jump matrix expressed in terms of 3-point solutions $\Psi_{\pm}\lb z\rb$. The latter will be considered as known, even though their explicit expressions in higher rank $N\ge 3$ are available only in a few special cases (rigid systems, etc).
  \end{enumerate}
  
  Let us now apply the results of Subsection~\ref{subsecwidom}. To the 4-point Fuchsian system (\ref{fuchs4}) we associate a tau function $\tau\lb t\rb\equiv \tau\left[J\right]$ defined by (\ref{adops})--(\ref{FR1c}) as a Fredholm determinant with a block integrable kernel given in terms of 3-point solutions. Theorem~\ref{WidomII} then yields
  \begin{cor}\label{cor4pt} Let $\tau_{\mathrm{JMU}}\lb t\rb$ denote the Jimbo-Miwa-Ueno tau function of (\ref{fuchs4}) defined by
  \beq\label{taujmu1}
  \partial_t\ln\tau_{\mathrm{JMU}}\lb t\rb=\frac{\operatorname{Tr}
  A_0 A_t}{t}+\frac{\operatorname{Tr}
    A_t A_1}{t-1}.
  \eeq  
  It coincides with $\tau\lb t\rb$ defined by (\ref{adops})--(\ref{FR1c}) up to a trivial prefactor,
  \beq\label{tau4tauJMU}
  \tau_{\mathrm{JMU}}\lb t\rb=\operatorname{const}\cdot t^{\frac12 \operatorname{Tr}\lb \mathfrak S^2-\Theta_0^2-\Theta_t^2\rb}
  \tau\lb t\rb.
  \eeq
  \end{cor}
  \pf On the circle $\mathcal C$, we have
  \beq\label{special01}
  \Psi_{\pm}\lb z\rb=\lb -z\rb^{-\mathfrak S}\Phi_{\pm}\lb z\rb,
  \qquad
  \bar\Psi_{\pm}\lb z\rb= {\Phi_{\mp}\lb z\rb}^{-1}\Phi\lb z\rb,
  \qquad
  J\lb z\rb={\Phi_{-}\lb z\rb}^{-1}\Phi_{+}\lb z\rb.
  \eeq
  It is convenient to choose the normalization of the auxiliary fundamental solutions so that $\Phi_+\lb z\rb\simeq \lb -z\rb^{\mathfrak S}$ as $z\to0$ and  $\Phi_-\lb z\rb\simeq \lb -z\rb^{\mathfrak S}$ as $z\to\infty$. In particular, in this normalization $\Phi_+\lb z\rb$ is independent of $t$. Substituting (\ref{special01}) into the
  Widom's formula (\ref{dertauRH}), we then obtain
  \beq\label{special02}
  \partial_t\ln\tau\lb t\rb=\frac1{2\pi i}\int_{\mathcal C}\operatorname{Tr}\lb \partial_t\Phi_-\,\Phi_-^{-1}\frac{\mathfrak S}{z} -\partial_t\Phi_-\,\Phi_-^{-1} \partial_z\Phi\;\Phi^{-1}\rb dz.
  \eeq
  Observe that the first and second term under trace are meromorphic, respectively, outside and inside $\mathcal C$ with the only possible pole at $z=\infty$ and $z=0,t$. This effectively reduces the above integral to residue calculation. Indeed, the analog of (\ref{fuchs4})
  for $\Phi_-$ is given by
    \beq \label{fuchs3}
    \partial_z\Phi_-=\Phi_- A^{-}\lb z\rb,\qquad A^-\lb z\rb=\frac{A^-_{0}}{z}+\frac{A^-_{t}}{z-t},
    \eeq
  where $A^-_0$, $A^-_t$ and $A^-_0+A^-_t$ belong to conjugacy classes of $\Theta_0$, $\Theta_t$ and $\mathfrak S$. Conservation of monodromy upon variation of $t$ gives  one more equation,
  $\partial_t \Phi_-=-\Phi_- A^-_t \lb z-t\rb^{-1}$, which implies that the first term in (\ref{special02})  (given by the residue at $z=\infty$) vanishes. The second term may be rewritten as
  \beq\label{special03}
  \frac1{2\pi i}\int_{\mathcal C}\operatorname{Tr}\lb  \frac{A^-_t}{z-t}\Phi_-^{-1}\Phi A\lb z\rb \Phi^{-1}\Phi_-\rb dz
  =\partial_t\ln\tau_{\mathrm{JMU}}\lb t\rb-\frac{\operatorname{Tr}A^-_0A^-_t}{t}+\operatorname{res}_{z=t}
  \frac{ \operatorname{Tr}A^-_t\Phi_-^{-1}\Phi A_t \Phi^{-1}\Phi_-}{\lb z-t\rb^2}.
  \eeq
  Here, the last expression  corresponds to the contribution with a 2nd order pole at $z=t$ and the first two are the residues of the rest at simple poles $z=t$ and $z=0$.
  Since $2\operatorname{Tr}A^-_0A^-_t=\operatorname{Tr}\lb \mathfrak S^2-\Theta_0^2-\Theta_t^2\rb$,  it now suffices to show that the last expression vanishes to finish the proof. Indeed, since 
  \ben\Phi\lb z\rb^{-1}\Phi_-\lb z\rb=G_{t}^{-1}\left[\mathbb 1+g_{t}\lb z-t\rb+O\lb \lb z-t\rb^2\rb\right] G_{t}^- \qquad
  \text{as } z\to t,
  \ebn
  with some $g_t\in\operatorname{Mat}_{N\times N}\lb \Cb\rb$, the last contribution to (\ref{special03}) is equal to $\operatorname{Tr}g_t\bigl[G_{t}^-A^-_t \lb {G_{t}^-}\rb^{-1},G_{t}A_t G_{t}^{-1}\bigr]$. But we have $G_{t}^-A^-_t \lb{G_{t}^-}\rb^{-1}=G_{t}A_t G_{t}^{-1}=\Theta_t$, and the statement follows.  \epf

  \subsubsection{Two irregular singularities}
  Let us now consider a linear system with two irregular singularities at $0$ and $\infty$ of respective Poincar\'e ranks ${R_0,R_{\infty}\in\Zb_{>0}}$. The general form of such a system reads
  \beq\label{IrrS}
  \partial_z\Phi=\Phi A\lb z\rb,\qquad A\lb z\rb=\sum_{k=-R_0}^{R_{\infty}}z^{k-1}A_{k},
  \eeq
  with $A_k\in\operatorname{Mat}_{N\times N}\lb \Cb\rb$ and $\operatorname{Tr} A_k=0$. The coefficients $A_{-R_{0}},A_{R_{\infty}}$ corresponding to the most singular terms at $z=0,\infty$, are assumed to be diagonalizable; we write $A_{-R_0}=G_0^{-1}\Theta^{\lb 0\rb}_{-R_{0}} G_0$, 
  $A_{R_{\infty}}=G_{\infty}^{-1}\Theta^{\lb \infty\rb}_{R_{\infty}} G_{\infty}$ with diagonal and traceless $\Theta^{\lb 0\rb}_{-R_{0}}$, $\Theta^{\lb \infty\rb}_{R_{\infty}}$. 
  
  There exist unique formal fundamental solutions
  \begin{align*}
  \Phi^{\lb a\rb}_{\mathrm{form}}\lb z\rb=e^{\Theta^{\lb a\rb}\lb z\rb}\hat{\Phi}^{(a)}\lb z\rb G_{a},\qquad a=0,\infty,
  \end{align*}
  with 
  \ben
  \begin{gathered}
  \hat{\Phi}^{\lb 0\rb}\lb z\rb=\mathbb 1+\sum_{k=1}^{\infty}g^{\lb 0\rb}_k z^k,\qquad
  \hat{\Phi}^{\lb \infty\rb}\lb z\rb=\mathbb 1+\sum_{k=1}^{\infty}g^{
  \lb \infty\rb}_k z^{-k}, \\
  \Theta^{\lb 0\rb}\lb z\rb=\sum_{k=-R_0}^{-1}\frac{\Theta^{\lb 0\rb}_k}{k}z^k+\Theta^{\lb 0\rb}_0\ln z,\qquad
  \Theta^{\lb \infty\rb}\lb z\rb=\sum_{k=1}^{R_{\infty}}\frac{\Theta^{\lb \infty\rb}_k}{k}z^k-\Theta^{\lb \infty\rb}_0\ln z,
  \end{gathered} 
  \ebn
  where all $\Theta^{\lb a\rb}_k$ are given by diagonal matrices. These matrices, together with the coefficients $g^{\lb a\rb}_k$, are uniquely fixed by the linear system \eqref{IrrS}. Genuine canonical solutions $\Phi^{\lb a\rb}_k\lb z\rb$ with $k=1,\ldots,2R_{a}+1$ are asymptotic to $\Phi^{\lb a\rb}_{\mathrm{form}}\lb z\rb$ in $2R_{a}+1$ Stokes sectors $\mathcal S^{\lb a\rb}_k$ around $z=a$, and are related by Stokes matrices $S_k=\Phi^{\lb a\rb}_{k+1}\lb z\rb{\Phi^{\lb a\rb}_k\lb z\rb}^{-1}$ on their overlap.  
  
   \begin{figure}[h!]
   \centering
   \includegraphics[height=7cm]{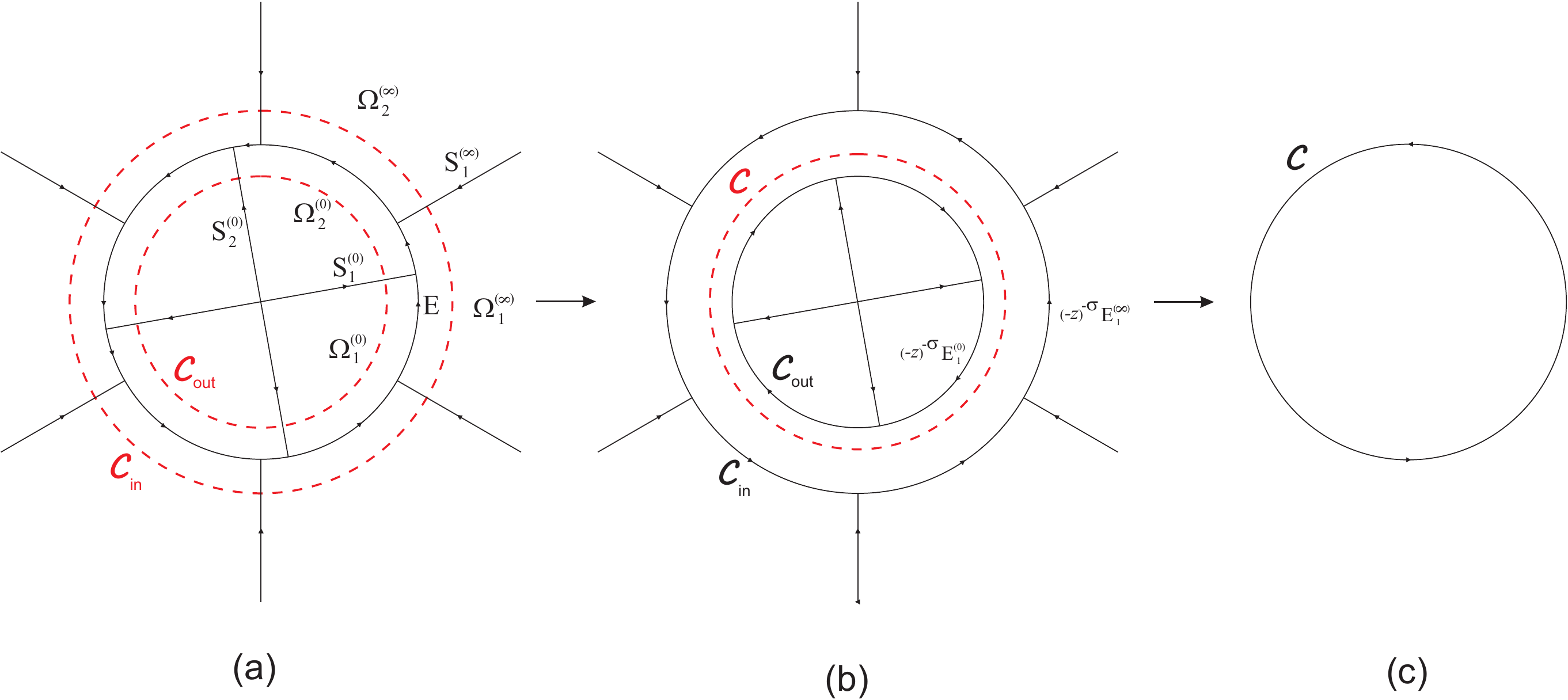}
   \begin{minipage}{0.65\textwidth}
   \caption{Transformation of RH contour for systems with two irregular singularities of Poincar\'e ranks $R_0=2$,  $R_{\infty}=3$.
                      \label{Toeplitz_Fig4}}
                                       \end{minipage}
                     \end{figure}

  The transformation of (\ref{IrrS}) into a Riemann-Hilbert problem on a circle is carried out similarly to the $4$-point Fuchsian case:
  \begin{enumerate}
  \item Introduce a function $\tilde{\Psi}\lb z\rb$ which coincides with the canonical solutions $\Phi^{\lb a\rb}_k\lb z\rb$ inside the sectors ${\Omega^{\lb a\rb}_k\subset\mathcal S^{\lb a\rb}_k}$ schematically represented in Fig.~\ref{Toeplitz_Fig4}a. The rays therein belong to overlaps of adjacent Stokes sectors. The function $\tilde{\Psi}\lb z\rb$ solves a dual RHP on the contour $\tilde\Gamma$, indicated in Fig.~\ref{Toeplitz_Fig4}a by solid black curves. Besides the jumps on the rays (given by the Stokes matrices), one has constant jumps on different arcs of the connection circle. All of the latter can be expressed in terms of one connection matrix, e.g. $E=\Phi^{\lb 0\rb}_{1}\lb z\rb{\Phi^{\lb \infty\rb}_1\lb z\rb}^{-1}$. There is also an asymptotic  condition
  $e^{-\Theta^{\lb a\rb}\lb z\rb}\tilde{\Psi}\lb z\rb=O\lb1\rb$ as $z\to a$ on~$\Cb\Pb^1\backslash\tilde{\Gamma}$.
  \item We would now like to cancel the jumps inside the open annulus $\hat{\mathcal A}$ bounded by the circles $\mathcal C_{\mathrm{out},\mathrm{in}}$ indicated in Fig.~\ref{Toeplitz_Fig4}a by dashed red lines. Pick any fundamental matrix solution $\tilde\Phi\lb z\rb$ of (\ref{IrrS}), e.g. $\Phi^{\lb 0\rb}_{1}\lb z\rb$. Let $M_0\in\mathrm{SL}\lb N,\Cb\rb$  be its anticlockwise monodromy around $z=0$. This matrix is determined up to conjugation by the Stokes matrices. Write $M_0=e^{2\pi i \mathfrak S}$ choosing $\mathfrak S$  so that $\operatorname{Tr}\mathfrak S=0$. Define
    \ben
    \hat\Psi\lb z\rb=\begin{cases}
      \lb -z\rb^{-\mathfrak S}\tilde\Phi\lb z\rb,\qquad& z\in\hat{\mathcal A},\\
      \tilde\Psi\lb z\rb, \qquad &z\notin\bar{\hat{\mathcal A}}.
      \end{cases}
    \ebn
  The dual RHP for $\hat\Psi\lb z\rb$ is posed on the contour $\hat{\Gamma}$ indicated in Fig.~\ref{Toeplitz_Fig4}b by solid black lines.
  \item As before in (\ref{psi1c}), it now suffices to divide $\hat\Psi\lb z\rb$ inside and outside of an auxiliary circle $\mathcal C$ by the solutions $\Psi_-\lb z\rb$ and $\Psi_+\lb z\rb$ of the auxiliary dual RHPs set on the two connected components $\hat \Gamma_-$ and $\hat\Gamma_+$ of $\hat{\Gamma}$, containing $\mathcal C_{\mathrm{out}}$
  and $\mathcal C_{\mathrm{in}}$. They are related to the solutions $\Phi_-\lb z\rb$ and $\Phi_+\lb z\rb$ of two auxiliary linear systems. The first one has an irregular singular point of Poincar\'e rank $R_0$ at $z=0$ and a regular singularity at $z=\infty$. In the second, there is a regular singular point at $z=0$ and an irregular singularity of Poincar\'e rank $R_{\infty}$ at $z=\infty$. We thereby obtain a dual RHP for a function $\bar{\Psi}\lb z\rb$, with the jump 
  \ben
  J\lb z\rb= {\Psi_-\lb z\rb}^{-1}\Psi_+\lb z\rb={\Phi_-\lb z\rb}^{-1}\Phi_+\lb z\rb,
  \ebn 
  on $\mathcal C$ written in the form of a direct factorization.
  Similarly to the above, we make the identifications
  $ \Psi_{\pm}\lb z\rb=\lb -z\rb^{-\mathfrak S}\Phi_{\pm}\lb z\rb$,
  $\bar\Psi_{\pm}\lb z\rb= {\Phi_{\mp}\lb z\rb}^{-1}\tilde\Phi\lb z\rb$.
  \end{enumerate}
 
   The set $\mathcal T$ of isomonodromic times  consists of the diagonal elements of 
   $\Theta^{\lb a\rb}_k$ with $k\ne 0$. We accordingly decompose it as $\mathcal T=\mathcal T^{\lb 0\rb}\cup \mathcal T^{\lb \infty\rb}$. 
 The Jimbo-Miwa-Ueno tau function of the system (\ref{IrrS}) is defined by the closed 1-form \cite[eq. (1.23)]{JMU}
   \beq\label{tauIrr}
   d_{\mathcal T}\ln\tau_{\mathrm{JMU}}\lb \mathcal T\rb=-\sum_{a=0,\infty}
   \operatorname{res}_{z=a}
   \operatorname{Tr}\lb \partial_z\hat{\Phi}^{\lb a\rb}\lb z\rb
   {\hat{\Phi}^{\lb a\rb}\lb z\rb}^{-1}
   d_{\mathcal T^{\lb a\rb}}\Theta^{\lb a\rb}\lb z\rb\rb.
   \eeq
  Let us now make contact between this formula and the construction given in Definition~\ref{tauRHPdef}.
  \begin{cor}
  Let $\tau^{\lb 0\rb}_{\mathrm{JMU}}\lb \mathcal T^{\lb 0\rb}\rb$ and $\tau^{\lb \infty\rb}_{\mathrm{JMU}}\lb \mathcal T^{\lb \infty\rb}\rb$ be the Jimbo-Miwa-Ueno tau functions of auxiliary linear systems for $\Phi_{-}\lb z\rb$ and $\Phi_+\lb z\rb$, and $\tau\left[J\right]$ be the Fredholm determinant defined by (\ref{adops})--(\ref{FR1c}). Then
  \beq\label{tau2irr}
  \tau \left[J\right] =\left[\tau^{\lb 0\rb}_{\mathrm{JMU}}\lb \mathcal T^{\lb 0\rb}\rb\tau^{\lb \infty\rb}_{\mathrm{JMU}}\lb \mathcal T^{\lb \infty\rb}\rb\right]^{-1} \tau_{\mathrm{JMU}}\lb \mathcal T\rb.
  \eeq
  \end{cor}
  \pf We choose again the normalization in which $\Phi_+\lb z\rb\simeq \lb -z\rb^{\mathfrak S}$ as $z\to0$ and  $\Phi_-\lb z\rb\simeq \lb -z\rb^{\mathfrak S}$ as $z\to\infty$. This implies that $\Phi_-\lb z\rb$ is independent of $\mathcal T^{\lb \infty\rb}$ and $\Phi_+\lb z\rb$ independent of $\mathcal T^{\lb 0\rb}$. From the Widom's differentiation formula then follows an analog of the equation (\ref{special01}),
    \beq\label{special11}
    d_{\mathcal T^{\lb 0\rb}}\ln\tau\left[J\right]=\frac1{2\pi i}\int_{\mathcal C}\operatorname{Tr}\lb d_{\mathcal T^{\lb 0\rb}}\Phi_-\,\Phi_-^{-1}\frac{\mathfrak S}{z} -d_{\mathcal T^{\lb 0\rb}}\Phi_-\,\Phi_-^{-1} \partial_z\tilde\Phi\;{\tilde\Phi}^{-1}\rb dz,
    \eeq
  and a similar formula for $d_{\mathcal T^{\lb \infty\rb}}\ln\tau\left[J\right]$. The first term in the integrand  of (\ref{special11}) is analytic outside $\mathcal C$ and the corresponding integral reduces to the residue at $z=\infty$. The isomonodromy equation for $\Phi_-$ has the form $d_{\mathcal T^{\lb 0\rb}}\Phi_-=\Phi_-U^-\lb z\rb$ with $U^-\lb z\rb=\sum_{k=-R_0}^{-1}z^{k} U_k^{-}$, which implies that this residue vanishes. The second term in the integrand extends to a meromorphic function inside $\mathcal C$ with the only possible pole at $z=0$, so that 
  \begin{align*}
  d_{\mathcal T^{\lb 0\rb}}\ln\tau\left[J\right]=&
  -\operatorname{res}_{z=0}\operatorname{Tr}\lb
  d_{\mathcal T^{\lb 0\rb}}\Phi_-\,\Phi_-^{-1} \partial_z\tilde\Phi\;{\tilde\Phi}^{-1}\rb=\\
  =&
  -\operatorname{res}_{z=0}\operatorname{Tr}\lb
    d_{\mathcal T^{\lb 0\rb}}\Phi_-\,\Phi_-^{-1} \partial_z\Theta^{
    \lb 0\rb}+ e^{-\Theta^{\lb 0\rb}}
    d_{\mathcal T^{\lb 0\rb}}\Phi_-\,\Phi_-^{-1}e^{\Theta^{\lb 0\rb}}\partial_z 
    \hat{\Phi}^{\lb 0\rb}{\hat{\Phi}^{\lb 0\rb}\,}^{-1}\rb=\\
    =&  -\operatorname{res}_{z=0}\operatorname{Tr}\lb
        d_{\mathcal T^{\lb 0\rb}}\Phi_-\,\Phi_-^{-1} \partial_z\Phi_- \Phi_-^{-1}-
        e^{-\Theta^{\lb 0\rb}}d_{\mathcal T^{\lb 0\rb}}\Phi_-\,\Phi_-^{-1} e^{\Theta^{\lb 0\rb}}\partial_z 
                \hat{\Phi}_-^{\lb 0\rb}{\hat{\Phi}_-^{\lb 0\rb}\,}^{-1}+ 
        d_{\mathcal T^{\lb 0\rb}}\Theta^{\lb 0\rb}\partial_z 
        \hat{\Phi}^{\lb 0\rb}{\hat{\Phi}^{\lb 0\rb}\,}^{-1}\rb=\\
        =&  -\operatorname{res}_{z=0}\operatorname{Tr}\lb
        U^-\lb z\rb A^-\lb z\rb
        - d_{\mathcal T^{\lb 0\rb}}\Theta^{\lb 0\rb}\partial_z 
                \hat{\Phi}^{\lb 0\rb}_-{\hat{\Phi}_-^{\lb 0\rb}\,}^{-1} +       
        d_{\mathcal T^{\lb 0\rb}}\Theta^{\lb 0\rb}\partial_z 
        \hat{\Phi}^{\lb 0\rb}{\hat{\Phi}^{\lb 0\rb}\,}^{-1}
        \rb,
  \end{align*}
  where $A^-\lb z\rb=\Phi_-^{-1}\partial_z\Phi_-$. Since $A^-\lb z\rb=
  \sum_{k=-R_0}^{0}z^{k-1} A_k^{-}$, the residue of the first term vanishes, while the second and third yield $-d_{\mathcal T^{\lb 0\rb}}\ln\tau^{\lb 0\rb}_{\mathrm{JMU}}+d_{\mathcal T^{\lb 0\rb}}\ln\tau_{\mathrm{JMU}}$. Similarly computing the differential $d_{\mathcal T^{\lb \infty\rb}}\ln\tau\left[J\right]$, we arrive at the expression (\ref{tau2irr}).
   \epf

  We have thus shown that $\tau\left[J\right]$ coincides with $\tau_{\mathrm{JMU}}\lb\mathcal T\rb$ up to more elementary factors  depending separately on $\mathcal T^{\lb 0\rb}$ and $\mathcal T^{\lb \infty\rb}$. These normalization factors are the tau functions of the auxiliary linear systems arising upon ``decorated pants decomposition'' of the Riemann sphere with 2 irregular punctures into two spheres with 1 irregular and 1 regular puncture. Schematically,
  \begin{subequations}
   \beq
   \begin{gathered}
   \tau_{\mathrm{JMU}}\lb\;
   \vcenter{\hbox{\includegraphics[height=5.5ex]{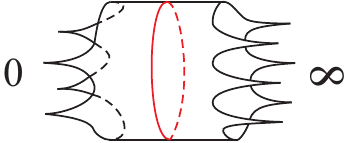}}}\;\rb= 
   \tau_{\mathrm{JMU}}\lb\;
     \vcenter{\hbox{\includegraphics[height=5.5ex]{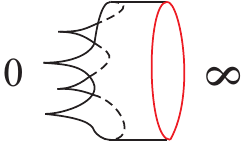}}}\;\rb
     \tau_{\mathrm{JMU}}\lb\;
      \vcenter{\hbox{\includegraphics[height=5.5ex]{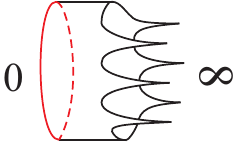}}}\;\rb\;
   \det\lb\begin{array}{cc}
   \mathbf 1 & \mathsf a\lb\;
     \vcenter{\hbox{\includegraphics[height=5ex]{CGLirrpantsR.eps}}}\;\rb
    \\ 
     \mathsf d\lb\;
        \vcenter{\hbox{\includegraphics[height=5ex]{CGLirrpantsL.eps}}}\;\rb & \mathbf 1
   \end{array}\rb .
     \end{gathered}
   \eeq
  The regular holes here correspond to Fuchsian singularities and cusps represent anti-Stokes directions.   
  The prefactor  $t^{\frac12 \operatorname{Tr}\lb \mathfrak S^2-\Theta_0^2-\Theta_t^2\rb}$ in (\ref{tau4tauJMU}) has a similar interpretation: it represents the isomonodromic tau function of the auxiliary Fuchsian system for $\Phi_-$ having singular points at $0$, $t$ and $\infty$, while the tau function of the auxiliary $3$-point system for $\Phi_+$ is just a constant, so that
     \beq
     \begin{gathered}
     \tau_{\mathrm{JMU}}\lb\;
     \vcenter{\hbox{\includegraphics[height=7ex]{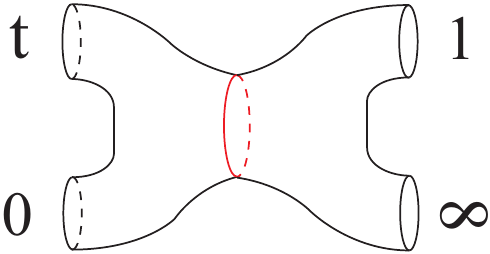}}}\;\rb= 
     \tau_{\mathrm{JMU}}\lb\;
       \vcenter{\hbox{\includegraphics[height=7ex]{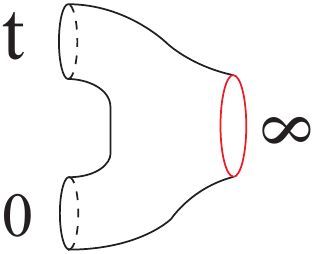}}}\;\rb
       \tau_{\mathrm{JMU}}\lb\;
        \vcenter{\hbox{\includegraphics[height=7ex]{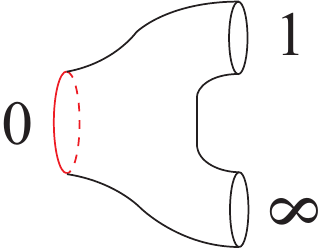}}}\;\rb\;
     \det\lb\begin{array}{cc}
     \mathbf 1 & \mathsf a\lb\;
       \vcenter{\hbox{\includegraphics[height=6ex]{CGLpantsR.eps}}}\;\rb
      \\ 
       \mathsf d\lb\;
          \vcenter{\hbox{\includegraphics[height=6ex]{CGLpantsL.eps}}}\;\rb & \mathbf 1
     \end{array}\rb .
       \end{gathered}
     \eeq
  \end{subequations}   
 The idea to associate Riemann surfaces with cusped boundaries to monodromy manifolds of isomonodromic systems in rank $N=2$ first appeared in \cite{CM,CMR}. Our results illustrate the use of the corresponding pictures in the analytic setting.
  
  \subsubsection{Integrable hierarchies\label{subsecIH}}
  As a second example, we will show how to apply our definition of tau function to the study of integrable hierarchies. The results outlined in this section are not new (see \cite{Cafasso,CafassoWu1,CafassoWu2,CDD}); the aim is to describe them in a way that makes the comparison with the case of isomonodromic deformations more transparent.
 To start with, consider a differential operator of fixed degree $N$ 
  \begin{equation}\label{L}
  	L := D^{N} + u_{N-2}D^{N-2} + \ldots + u_0,
  \end{equation}
  where we denoted $D := \partial_x$, and the coefficients $u_0, \ldots, u_{N - 2}$ depend on $x$ and some additional parameters we are now going to describe. The isospectral deformations of $L$ are described by the Lax system
  \begin{equation}\label{LaxSystem}
  \left\{
  	\begin{array}{rcl}
  	L\, \phi &=& z \phi,\\
  	\partial_{ t_j}\phi &=& \lb L^{j/N}\rb_+\phi, \qquad j \ne 0\;\mathrm{mod}\; N .
  \end{array} 
  \right.
  \end{equation}
  giving rise to the Gelfand-Dickey equations for the variables $\left\{u_0\lb x,\bt\rb, \ldots, u_{N - 2}\lb x,\bt\rb\right\}$, written in the Lax form as the compatibility conditions of the system \eqref{LaxSystem}:
  \begin{equation}
  \partial_{t_j} L = \left[\lb L^{j/N}\rb_+,L\right], \qquad j \ne 0\;\mathrm{mod}\; N .
  \end{equation}
 Here and below $\bt$ denotes the collection of all the deformation parameters $\bt := \left\{t_1,\ldots,t_{N-1},t_{N+1},\ldots \right\}$. 
 
 Converting the equations \eqref{LaxSystem} into a 1st order system of size $N$, we get
  \begin{equation}\label{LaxSystem2}
  \left\{
  	\begin{array}{rcl}
  \partial_x \Phi &=& \Phi \left(\begin{array}{ccccc}
  													0 & 0 & \ldots & 0 & z - u_0\\
  													1 & 0 & \ldots & 0 & -u_1 \\
  													0 & 1 & \ldots & 0 & -u_2\\
  													\vdots & 0 & \ddots & 0 & \vdots\\
  													0 & \ldots & \ldots & 1 & 0
												\end{array}\right),\\
  	\partial_{t_j}\Phi &=& \Phi M_j, \qquad j \ne 0\;\mathrm{mod}\; N.
  \end{array} 
  \right.
  \end{equation}
  where the matrices $M_j$ are completely determined by the coefficients of $\lb L^{j/N}\rb_+$. We fix uniquely a fundamental solution $\Phi\lb x,\bt;z\rb$ by
   imposing the asymptotics at infinity
   \beq\label{asympGD}
   			\Phi\lb x,\bt;z\rb = e^{x\Lambda + \sum_{j \ne 0\;\mathrm{mod}\; N}t_j \Lambda^j}\left[\mathbb 1 + O\lb z^{-1}\rb \right] = e^{x\Lambda + \sum_{j \ne 0\;\mathrm{mod}\; N}t_j \Lambda^j}\Phi_-\lb x,\bt;z\rb, \qquad z\to \infty ,
   \eeq
  where 
  \ben
  	\Lambda := \left(\begin{array}{ccccc}
  													0 & 0 & \ldots & 0 & z \\
  													1 & 0 & \ldots & 0 & 0 \\
  													0 & 1 & \ldots & 0 & 0\\
  													\vdots & 0 & \ddots & 0 & \vdots\\
  													0 & \ldots & \ldots & 1 & 0
												\end{array}\right).
  \ebn
  Indeed, in some cases $\Phi$ will be a genuine analytic function in a neighborhood of $\mathcal C$, and the corresponding solution belongs to the Segal--Wilson Grassmannian. Otherwise, one can still consider $\Phi$ just as a formal series, and in this case the solution belongs to the Sato Grassmannian.
  We now define our direct RHP by imposing
  \beq\label{JumpGD}
  	J\lb x,\bt;z\rb := \Phi^{-1}_-\lb 0,0;z\rb e^{-x\Lambda - \sum_{j \ne 0\;\mathrm{mod}\; N}t_j \Lambda^j}  =: \Psi_-^{-1}\lb z\rb\Psi_+\lb x,\bt;z\rb. 
  \eeq
In \cite{Cafasso,CafassoWu1} it has been proven that the related tau function defined in Definition \ref{tauRHPdef} coincides with the one defined by Segal and Wilson.
Indeed, one can act with the matrix-valued series $\Psi_-\lb z\rb$ on the subspace $H_+$, thus obtaining a subspace $W := H_+\cdot\Psi_- $ in $\mathrm{Gr}\lb H\rb$. The operator $-\mathsf{d} : H_{_+} \to H_-$ is nothing but the operator whose graph gives the subspace $W$ (i.e. the operator $A$ in the notation of \cite{SW}) and the formulas to be compared are \cite[eq. (3.5)]{SW} and
$$\tau[J] = \mathrm{det}_H\lb \mathbb{1} + L\rb = \mathrm{det}_{H_+}\lb \mathbb 1 - \mathsf{a}\mathsf{d}\rb.$$

Concerning the combinatorial expansion in the Subsection \ref{subseccombi}, we start by observing that the matrix $G^{[+]}$ does not depend on the particular solution to be studied, and can be computed explicitly. It reads
\beq\label{SchurMatrix}
	G^{[+]} = \left(\begin{array}{cccc|cccc}
				\ddots & \vdots & \vdots & \vdots & \vdots & \vdots & \vdots & \vdots \\
				\ldots  & 1 & s_1 & s_2 & s_3 & s_4 & s_5 & \ldots \\
				\ldots &  0 & 1 & s_1 & s_2 & s_3 & s_4 & \ldots \\
				\ldots  & 0 & 0 & 1 & s_1 & s_2 & s_3 & \ldots
			\end{array}\right),
\eeq
 where the elementary Schur polynomials $s_j = s_j\lb \bt\rb$ are defined by the relation
 \beq
 	\sum_{j \geq 0} s_j(\bt)z^j = \exp\left( \sum_{j \neq 0 \, \mathrm{mod}\, N} t_j z^j \right).
 \eeq
 In particular, its minors give, by definition (see for instance \cite{Macdonald}), the Schur polynomials associated to an arbitrary partition. Equivalently, one can compute the same Schur polynomial through the principal minor of $\mathsf{a}$, and the equivalence between the two approaches is given by the Giambelli identity \cite{EH}. As we mentioned before, the graph of the function $-\mathsf{d}$ determines the point $W$ associated to the particular solution we want to study. This means that its minors give the Pl\"ucker coordinates of $W$, which are equivalently computed as the minors of $G^{[-]}$, again via the Giambelli identity. Hence, the combinatorial formula \eqref{SatoTau},  in this case, is the standard tau function expansion in terms of Schur polynomials and Pl\"ucker coordinates.
 
 Note that, in this case, \emph{any} loop $\Psi_-\lb z\rb$ with the prescribed asymptotics $\mathbb{1} + O\lb z^{-1}\rb$ determines a solution of the hierarchy, so that it does not make much sense to ask about the general form of $-\mathsf{d}$. On the other hand, one can wonder how the already known solutions of the Gelfand-Dickey equations are obtained  through our procedure, and this had been answered already for quite a few families:
 \begin{itemize}
 	\item Suppose that $\Psi_-(z) = e^{X\lb z\rb}$, with $X(z)$ a nilpotent element of $\mathfrak{sl}_N[z^{-1}]$. Then there is just a finite number of non-zero minors of $\mathsf{d}$. In this case, the combinatorial expansion \eqref{SatoTau} becomes a finite sum and the tau function will be a polynomial obtained by particular linear combinations of Schur polynomials, see \cite{CDD}. These solutions are much studied in the literature, as their zeros evolve according to the (rational) Calogero--Moser hierarchy (see \cite{Wilson} and references therein).
	\item More generally, if $\Psi_-(z)$ has a truncated expansion (i.e. just a finite number of non-zero Fourier coefficients, say up to $n$) then, using a result obtained by Widom \cite{Widom1}, one can compute the Szeg\H o-Widom constant of $J^{-1}$ as the (finite-size) determinant $\det T_n\left[J\right]$. Such solutions are the ones associated to rational curves with singularities (see \cite{SW}), i.e. they correspond to multi--solitons, see \cite{Cafasso}.
	\item  Consider a Riemann surface of type (symmetric $N$-covering)
	\beq\label{RiemannSurface}
		\lambda^N = \prod_{j = 1}^{Nk +1} \lb z - a_j\rb = P\lb z\rb,
	\eeq
	and define $N$ functions $\left\{w_1\lb z\rb, \ldots , w_N\lb z\rb \right\}$ by
	\beq
		w_i\lb z\rb := \left(\frac{P\lb z\rb}{z}\right)^{\frac{i-1}N} \frac{1}{\prod_{j = 1}^{(i-1)k}\lb z-a_j\rb}, \quad i = 1,\ldots,N.
	\eeq
	The point in the Grassmannian generated by
     $\Psi_-\lb z\rb := \mathrm{diag}\lb w_1\lb z\rb,\ldots,w_N\lb z\rb\rb$
	belongs to the so-called Krichever locus \cite{SW}. It corresponds to an algebro-geometric solution of the Gelfand-Dickey hiearchy, whose spectral curve is exactly  \eqref{RiemannSurface}. The Schur expansion for these tau functions had been studied in great detail in \cite{EH}. A way to compute the tau function in this case is using the Widom's differentiation formula \eqref{dertauRH}. Indeed, here one can reduce the dual RHP \eqref{RHPdual} to a RHP with \emph{constant} jumps on the intervals $[a_1,a_2] \cup \ldots \cup [a_{Nk-1},a_{Nk}] \cup [a_{Nk+1},\infty)$, and solve it explicitly in terms of theta functions associated to the curve \eqref{RiemannSurface}, as described in \cite{Cafasso} for the case $N = 2$ (i.e. for hyper-elliptic curves). The procedure is a classical one, used already in the 80s by the Saint Petersburg school in the context of the so-called finite-gap integration (see for instance \cite{Matv} and references therein). The idea to use it to compute the Szeg\H o-Widom constant associated to matrix-valued algebraic symbols had been developed for the first time in \cite{IJK,IMM}. 
	\item As an example of solutions not belonging to the Segal-Wilson Grassmannian, we consider \emph{topological} solutions, uniquely fixed by the equations of the hierarchy and the additional string equation
	\beq
		\left(\sum_{i \neq 0\,\mathrm{mod}\, N }\left( \frac{i+ N}{N} t_{i+
N}-\delta_{i,1}\right)\frac{\partial }{\partial t_{i}}+\frac{1}{2 N}\sum_{ i+j= N}i j t_i t_j\right)\tau({\mathbf t}) =0.
	\eeq
	For $N = 2$, this is the celebrated Witten-Kontsevich tau function \cite{Kon}, and for $N > 2$ this is the Frobenius potential, to all genera, of the singularity of type $A_{N-1}$. Finding the element $\Psi_-\lb z\rb$ defining the relevant jump \eqref{JumpGD} corresponds to finding the point in the Sato-Segal-Wilson Grassmannian associated to the given solution of the hierarchy. This had been achieved, in the 90s, by Kac and Schwarz \cite{KS}, who proved that $W =H \cdot \Psi_-$ is uniquely fixed by the two conditions\footnote{Note that here we are working on the vector-valued Grassmannian, while in the cited paper the two conditions are (equivalently) stated in the scalar Grassmannian and the Kac-Schwarz operator, in particular, is a scalar one.}
	\beq\label{KacSchwarz}
		zW \subseteq W, \quad \mathcal R_N W \subseteq W,
	\eeq
	where $\mathcal R_N$ is the differential operator
	\beq
		\mathcal R_N = \partial_z - \Lambda + \lb N z\rb^{-1}\rho,
	\eeq
and $\rho$ is a diagonal traceless matrix whose coefficients depend on the Cartan matrix of the Lie algebra $\mathfrak{sl}_N$, see for instance \cite[eq. (3.24)]{CafassoWu2}. Another way of stating the second equation in \eqref{KacSchwarz} is to say that $\Psi_-\lb z\rb$ satisfies the so-called \emph{reduced string equation}
	\beq\label{reducedSE}
		\mathcal R_N\Psi_- = \Psi_-\lb\Psi_-^{-1}\Lambda\Psi_-\rb_+,
	\eeq
	and it had been proven in \cite{CafassoWu2} that there exists a unique solution $\Psi_-(z) = {\rm e}^{X(z)}$, with $X\lb z\rb$ an element in the loop (sub)algebra $\mathfrak{sl}_N[[z^{-1}]]$ . This solution is in general just a formal series, exactly as the corresponding tau function, whose coefficients give intersection numbers on the Deligne-Mumford moduli space of stable curves. Note that \eqref{reducedSE} is of the same form as \eqref{DEcond}. This is not surprising, as the connection between isomonodromic deformations and string equations goes back to the work \cite{Moore} (see also \cite{ACvM}, where this connection is established using the Kac-Schwarz operators described above).
 \end{itemize}
 In order to simplify the notations, in this subsection we only considered the case of Gelfand-Dickey hierarchies. Nevertheless, most of the results described above apply as well to the more general case of the Drinfeld-Sokolov hierarchies associated to arbitrary (affine) Kac-Moody algebras \cite{DS}. The idea is to consider direct RHPs as the one in \eqref{JumpGD} above, but with $\Psi_-$ an element of the form $\Psi_-\lb z\rb = {\rm e}^{X\lb z\rb}$, with $X\lb z\rb \in \mathfrak{g}[[z^{-1}]]$, and $\mathfrak{g}$ an arbitrary (finite-dimensional) Lie algebra. The element $\Psi_+$, on the other hand, will be replaced by  $$\Psi_+\lb x,\mathbf{t};z\rb = {\rm e}^{-x\Lambda_1 - \sum_{j \in E_+}t_j\Lambda_j},$$ where $E_+$ is the set of the (positive) exponents of the Kac-Moody algebra $\mathfrak g[z,z^{-1}] \oplus \mathbb C c$, and $\left\{\Lambda_j,\, j \in E_+ \right\}$ is (half of) the Heisenberg sub-algebra associated to an arbitrary gradation of the algebra. Polynomial and topological solutions of these hierarchies had been treated, using this formalism, in \cite{CDD, CafassoWu2}. It would be interesting (but technically involved, because of the size of matrix representations) to study algebro-geometric solutions associated to arbitrary Drinfeld-Sokolov hierarchies.

  \section{General contour}\label{secmulti}
  The Definition~\ref{tauRHPdef} of $\tau\left[J\right]$ makes sense if we replace the circle $\mathcal C$  by a finite collection $\Gamma=\bigcup_{a=1}^M\mathcal C_a$ of non-intersecting smooth closed curves which we sometimes call ovals. However, defining the jump of the relevant dual RHP in the form of direct factorization is no longer natural from the point of view of applications, which makes the Fredholm determinant representation (\ref{FR1c}) and the differentiation formula (\ref{dertauRH}) less useful in this setting. What we would like to have instead are the formulae of the same type but expressed in terms of the direct factorization of the \textit{individual} jumps on each curve~$\mathcal C_a$. 
  
\begin{figure}[h!]
   \centering
   \includegraphics[height=6cm]{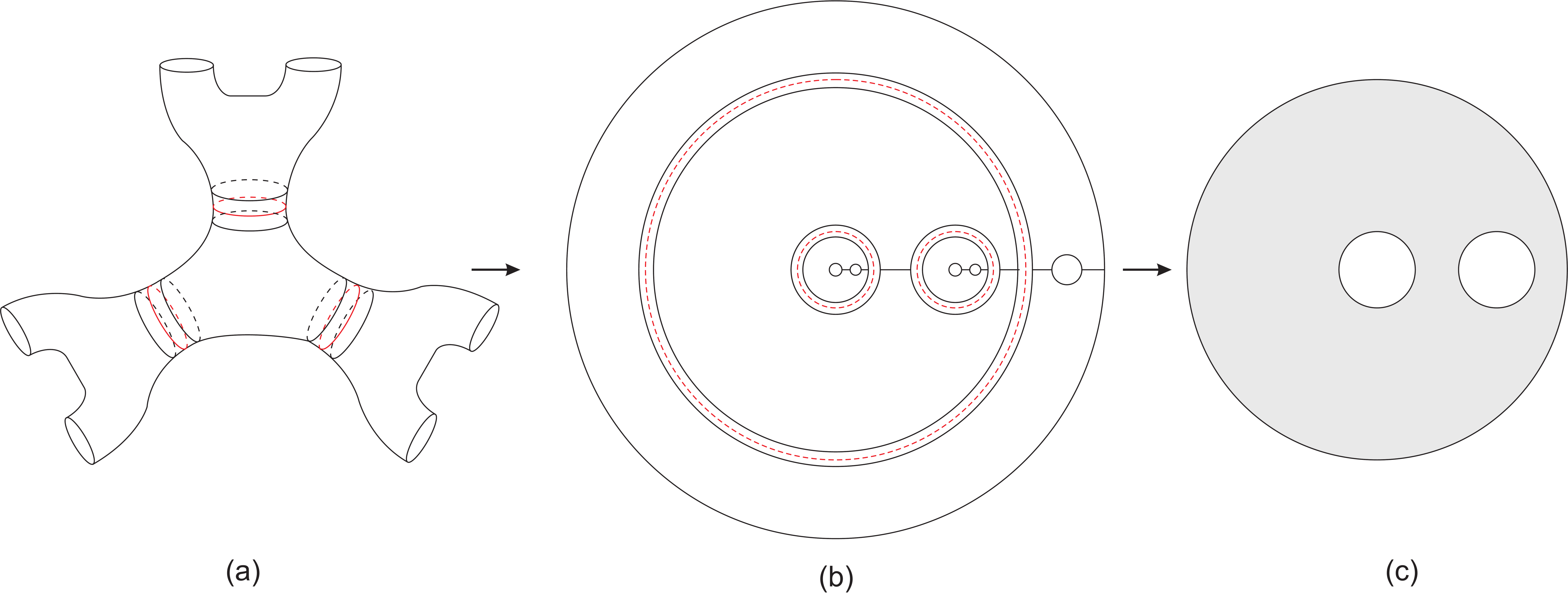}
   \begin{minipage}{0.65\textwidth}
   \caption{``Sicilian'' RH contour for $6$-point Fuchsian systems.
                      \label{fig_sic}}
                                       \end{minipage}
                     \end{figure}

  The existence of such expressions is suggested by the recent work \cite{GL16} by two of the authors, where Fredholm determinant and combinatorial series representations were obtained for the tau function of the $n$-point Fuchsian system --- including, in particular, the tau function of the Garnier system $\mathcal G_{n-3}$. The paper \cite{GL16} deals with the special case where the contour is given by a collection of concentric circles coming from a ``linear'' pants decomposition of $\Cb\Pb^1\backslash\left\{n\text{ points}\right\}$. Our aim here is to extend these results to RHPs with more general jumps on arbitrary configurations of ovals such as the one represented in Fig.~\ref{fig_sic}c.

    \subsection{Notations and setting of the RHP\label{subsecnotationmany}}    
    The complement of $\Gamma$ in $\Cb\Pb^1$ has $M+1$ connected components, which will be called faces (or pants). It admits a unique, up to permutation, 2-coloring by colors $\left\{+,-\right\}$ which will be fixed once and for all. Denote by $\mathsf F_{\pm}$ the set of faces of color $\pm$, by $\mathsf C:=\left\{\mathcal C_1,\ldots, \mathcal C_M\right\}$ the set of all ovals and by $\mathsf C_f$ the set of boundary curves of the face $f$. Let $\phi_{\pm}:\mathsf C\to\mathsf F_{\pm}$ be the map assigning to each curve $\mathcal C\in\mathsf C$ the unique face in $\mathsf F_{\pm}$ having $\mathcal C$ among the boundary components.

    The coloring allows to choose a convenient orientation of $\Gamma$ for which all faces of color $+$ (or $-$) are located on the positive (resp. negative) side of their boundary curves. We denote by $\varphi_+\lb \mathcal C\rb$ and  $\varphi_-\lb \mathcal C\rb$ the closure of interior and exterior of the curve $\mathcal C$ with respect to the above orientation. Clearly $\phi_{\pm}\lb \mathcal C\rb\subseteq\varphi_+\lb \mathcal C\rb$, but in general $\varphi_\pm\lb \mathcal C\rb$ can contain faces of different colors.
   Assign to every boundary $\mathcal C\in\mathsf C$ a pair of functions $\Psi_{\mathcal C,\pm}:\mathcal C\to\mathrm{GL}\lb N,\Cb\rb$ that continue analytically to $\varphi_{\pm}\lb \mathcal C\rb$. Also, to every $\mathcal C\in \mathsf C$ we assign the space of functions $H_{\mathcal C}=L^2\lb \mathcal C,\Cb^N\rb$. It may be decomposed as $H_{\mathcal C}=H_{\mathcal C,+}\oplus H_{\mathcal C,-}$, where $H_{\mathcal C,\pm}$ consist of functions that continue to $\varphi_{\pm}\lb \mathcal C\rb$ (and vanish at $\infty$ on the appropriate side of $\mathcal C$). This means in particular that the columns of $\Psi_{\mathcal C,-}$ do not necessarily belong to $H_{\mathcal C,-}$ but e.g. $\Psi_{\mathcal C,-}H_{\mathcal C,-}\subseteq H_{\mathcal C,-}$. \vspace{0.2cm}\\   
   \textit{Riemann-Hilbert problem}. We wish to consider the dual RHP posed  on $\Gamma=\bigcup_{a=1}^M \mathcal C_a$ in which the jumps are given in the form of direct factorization,
   \ben
    J_{\mathcal C}\equiv J\bigl|_{\mathcal C}:=\Psi_{\mathcal C,-}^{-1}
    \Psi_{\mathcal C,+},\qquad \mathcal C\in\mathsf C.
   \ebn
   That is, we want to find an analytic invertible matrix function $\bar\Psi\lb z\rb$ on $\Cb\Pb^1\backslash\Gamma$ whose boundary values on  $\Gamma$ satisfy  $J=\bar\Psi_+{\bar\Psi_-}^{-1}$.\vspace{0.2cm}
   
   Denote by $\Pi_{\mathcal C,\pm}$ the projections on $H_{\mathcal C,\pm}$ along $H_{\mathcal C,\mp}$ and consider
   \ben
   H=H_+\oplus H_-,\qquad H_{\pm}=\bigoplus_{\mathcal C\in\mathsf C}H_{\mathcal C,\pm}.
   \ebn
   For $\mathcal C,\mathcal C'\in\mathsf C$, $\mathcal C\in\varphi_{\pm}\lb\mathcal C'\rb$,  we define the operators $\Pi_{ \mathcal C\leftarrow\mathcal C',\pm}: H_{\mathcal C'}\to H_{\mathcal C}$ such that $\Pi_{ \mathcal C\leftarrow\mathcal C',\pm}g$ is the restriction to $\mathcal C$ of the analytic continuation of $\Pi_{\mathcal C',\pm} g$ to  $\varphi_{\pm}\lb \mathcal C'\rb$. In particular, for $\mathcal C\in \mathsf C_{\phi_{\pm}\lb\mathcal C'\rb}$, $\mathcal C\ne\mathcal C'$ we have $\operatorname{im}\lb \Pi_{ \mathcal C\leftarrow\mathcal C',\pm }\rb\subseteq H_{\mathcal C,\mp}$, whereas for $\mathcal C=\mathcal C'$ one has $\Pi_{ \mathcal C'\leftarrow\mathcal C',\pm}=\Pi_{\mathcal C',\pm}$.
   \begin{defin}\label{defmultitau}
   The tau function $\tau\left[J\right]$ associated to the above RHP is defined as the Fredholm determinant
   \beq\label{taugen}
   \tau\left[J\right]=\operatorname{det}_H\lb \mathbb 1+L \rb, \qquad 
   L=\lb\begin{array}{cc} 0 & \mathsf A_{+-} \\ \mathsf A_{-+} & 0\end{array}\rb
   \in\operatorname{End}\lb H_+\oplus H_-\rb,
   \eeq
   where the operators $\mathsf A_{\pm\mp}:H_\mp\to H_\pm$ are defined by
   \begin{align*}
   \mathsf A_{\pm\mp}=&\,\sum_{f\in\mathsf F_{\mp}}\sum_{\mathcal C,\mathcal C'\in\mathsf C_f}\mathsf A_{\mathcal C,\pm;\mathcal C',\mp},\\
   \mathsf A_{\mathcal C,\pm;\mathcal C',\mp}=&\, 
   \Psi_{\mathcal C,\pm}
   \Pi_{ \mathcal C\leftarrow\mathcal C',\mp}\Psi_{\mathcal C',\pm}^{\;-1}-\delta_{\mathcal C,\mathcal C'}\;\Pi_{\mathcal C',\mp}.
   \end{align*}
   \end{defin}

   One can consider elementary summands $\mathsf A_{\mathcal C,\pm;\mathcal C',\mp}$ as integral operators on $H$,
   \ben
   \lb\mathsf A_{\mathcal C,\pm;\mathcal C',\mp} g\rb\lb z\rb=
   \frac1{2\pi i}\oint_{\mathcal C'}\mathsf A_{\mathcal C,\pm;\mathcal C',\mp}\lb z,z'\rb g\lb z'\rb dz',
   \ebn   
   whose integral kernels have integrable form and are given by
   \ben
   \mathsf A_{\mathcal C,\pm;\mathcal C'\mp}\lb z,z'\rb
   =\pm\chi_{\mathcal C}\lb z\rb \frac{\Psi_{\mathcal C,\pm}\lb z\rb {\Psi_{\mathcal C',\pm}\lb z'\rb}^{-1}-\mathbb 1\;\delta_{\mathcal C,\mathcal C'}}{z-z'}\chi_{\mathcal C'}\lb z'\rb,
   \ebn
   with $\chi_{\mathcal C}\lb z\rb$ denoting the indicator function of $\mathcal C$. We leave it as an exercise for the reader to verify that $\operatorname{im}\mathsf A_{\mathcal C,\pm;\mathcal C',\mp}\subseteq H_{\pm}\subseteq \operatorname{ker}\mathsf A_{\mathcal C,\pm;\mathcal C',\mp}$.     The above expressions provide a many-oval counterpart of the one-circle formula \eqref{FR1c}. The operators $\mathsf A_{+-}$ and $\mathsf A_{-+}$ are analogs of the operators $\mathsf a$ and~$\mathsf d$. They can be regarded as  matrices whose operator entries are labeled by pairs of curves in $\mathsf C$; these entries are non-zero only if the relevant curves bound the same face.
 
   \begin{rmk}
  One may wonder whether it is also possible to obtain an analog of the Widom's determinant (\ref{tauRHP}), i.e. not to use direct factorization and express $\tau\left[J\right]$ solely in terms of the jumps.
     The answer is positive and can be obtained by conjugation of $L$ by the multiplication operator $\lb\bigoplus_{\mathcal C\in\mathsf C}\Psi_{\mathcal C,+}\rb\oplus \lb\bigoplus_{\mathcal C\in\mathsf C}\Psi_{\mathcal C,-}\rb$: one can equivalently write
     \beq\label{taugen2}
     \begin{gathered}
     \tau\left[J\right]=\operatorname{det}_H\lb \mathbb 1+\tilde L \rb, \qquad 
     \tilde L=\lb\begin{array}{cc} 0 & \tilde{\mathsf A}_{+-} \\ \tilde{\mathsf A}_{-+} & 0\end{array}\rb,\\
     \tilde{\mathsf A}_{\pm\mp}=\sum_{f\in\mathsf F_{\mp}}\sum_{\mathcal C,\mathcal C'\in\mathsf C_f}\lb
     \Pi_{ \mathcal C\leftarrow\mathcal C',\mp}J_{\mathcal C'}^{\mp1}- \delta_{\mathcal C,\mathcal C'}\;J_{\mathcal C'}^{\mp 1}\rb\Pi_{\mathcal C',\mp}.
     \end{gathered}
     \eeq
   \end{rmk}

  \subsection{Differentiation formula\label{subsec43}}
  Let us now establish the many-oval counterpart of the differentiation formula given in Theorem~\ref{WidomII}. The first step is the calculation of the inverse $\lb\mathbb 1+L\rb^{-1}$. To this end we first rewrite $\mathbb 1+L$ in the form
  \beq\label{rhs1}
  \mathbb 1+L=\lb\begin{array}{cc}\sum\limits_{f\in\mathsf F_{+}}\sum\limits_{\mathcal C,\mathcal C'\in\mathsf C_f}
  \Psi_{\mathcal C,-}
     \Pi_{ \mathcal C\leftarrow\mathcal C',+}\Psi_{\mathcal C',-}^{\;-1} &
  \sum\limits_{f\in\mathsf F_{-}}\sum\limits_{\mathcal C,\mathcal C'\in\mathsf C_f}
    \Psi_{\mathcal C,+}
       \Pi_{ \mathcal C\leftarrow\mathcal C',-}\Psi_{\mathcal C',+}^{\;-1}   
     \end{array}\rb,
  \eeq
  where the first and second column correspond to the action of
  $\mathbb 1+L$ on $H_+$ and $H_-$. Let us note that 
  \ben
  \Gamma=\bigcup\nolimits_{f\in\mathsf F_\pm}\bigcup\nolimits_{\mathcal C\in\mathsf C_f}\mathcal C.
  \ebn 
  It is useful to interpret the contributions $\mathcal P_{\oplus,\pm}^{[f]}:= \sum\limits_{\mathcal C,\mathcal C'\in\mathsf C_f}
      \Psi_{\mathcal C,\mp}
         \Pi_{ \mathcal C\leftarrow\mathcal C',\pm}\Psi_{\mathcal C',\mp}^{\;-1}$ of individual faces $f\in\mathsf F_{\pm}$ to the above sums as integral operators acting from $\bigoplus\limits_{\mathcal C\in\mathsf C_f}H_\pm\lb\mathcal C\rb$ to $\bigoplus\limits_{\mathcal C\in\mathsf C_f}H\lb\mathcal C\rb$ by
  \beq\label{rhs2}
  \lb\mathcal P_{\oplus,\pm}^{[f]} g^{[f]}\rb\lb z\rb=\pm
 \sum_{\mathcal C,\mathcal C'\in\mathsf C_f} \frac{1}{2\pi i}\oint_{\mathcal C'}  \frac{
 \chi_{\mathcal C}\lb z\rb\Psi_{\mathcal C,\mp}\lb z\rb
  \Psi_{\mathcal C',\mp}\lb z'\rb^{\;-1} 
  g^{[f]}_{\mathcal C'}\lb z'\rb dz'}{z'-z}.
  \eeq
  The contour of integration is deformed to the face $\phi_{\mp}\lb \mathcal C\rb$ (i.e. outside the face $f$) whenever it becomes necessary to interpret the singular factor $1/\lb z'-z\rb$. 
  
  Next we construct in a similar fashion the operators
  $\mathcal P_{\Sigma,\pm}:H\to H_{\pm}$ defined by
  \beq\label{psigpm}
  \lb\mathcal P_{\Sigma,\pm} g\rb\lb z\rb=
  \pm
   \sum_{\mathcal C,\mathcal C'\in\mathsf C}\Pi_{\mathcal C,\pm} \frac{1}{2\pi i}\oint_{\mathcal C'}  \frac{
   \chi_{\mathcal C}\lb z\rb\Psi_{\mathcal C,+}\lb z\rb\bar\Psi_{\mathcal C,-}\lb z\rb
          \bar\Psi_{\mathcal C',-}\lb z'\rb^{\;-1}\Psi_{\mathcal C',+}\lb z'\rb^{-1} 
    g_{\mathcal C'}\lb z'\rb dz'}{z'-z}.
  \eeq
  The convention for the contour is the same, i.e. it is pushed away slightly to the negative (positive) faces for $\mathcal P_{\Sigma,+}$
  (resp. $\mathcal P_{\Sigma,-}$). In contrast to the face operators $\mathcal P_{\oplus,\pm}^{[f]}$, the operators $\mathcal P_{\Sigma,\pm}$ involve the solution $\bar\Psi_{\mp}$ of the dual RHP. Constructing this solution is essentially equivalent to the calculation of the resolvent $\lb \mathbb 1+L\rb^{-1}$ thanks to the following lemma.
  
  \begin{lemma} If the dual RHP is solvable, then 
  $\lb\mathbb 1+L\rb^{-1}=
    \ds\lb\begin{array}{c} \mathcal P_{\Sigma,+} \\ \mathcal P_{\Sigma,-}\end{array}\rb$.
  \end{lemma}
  \pf Let us compute, for instance, the action of the ``$+-$'' component of the product $\ds\lb\begin{array}{c} \mathcal P_{\Sigma,+} \\ \mathcal P_{\Sigma,-}\end{array}\rb \lb \mathbb 1+L\rb$. Given $g_-\in H_-$, it reads
  \begin{align*}
  &\Bigl (\mathcal P_{\Sigma,+}\sum\limits_{f\in\mathsf F_{-}}\sum\limits_{\mathcal C,\mathcal C'\in\mathsf C_f}
      \Psi_{\mathcal C,+}
         \Pi_{ \mathcal C\leftarrow\mathcal C',-}\Psi_{\mathcal C',+}^{\;-1} g_-\Bigr)\lb z\rb=\\&=-\!\!\!
         \sum_{\mathcal C,\mathcal C''\in\mathsf C}\sum_{\;\;\mathcal C'\in\mathsf C_{\phi_-\lb \mathcal C''\rb}}\!\!\!\Pi_{\mathcal C,+}\frac{1}{\lb 2\pi i\rb^2}
    \oint_{\mathcal C''}     \oint_{\mathcal C'}
   \frac{
      \chi_{\mathcal C}\lb z\rb\Psi_{\mathcal C,+}\lb z\rb\bar\Psi_{\mathcal C,-}\lb z\rb
             \bar\Psi_{\mathcal C',-}\lb z'\rb^{\;-1}\Psi_{\mathcal C'',+}\lb z''\rb^{-1} 
       g_{\mathcal C'',-}\lb z''\rb dz'dz''}{\lb z'-z\rb \lb z'' -z'\rb}.      
  \end{align*}
  Recall that in this expression, the contour of integration w.r.t. $z''$ is slightly deformed to the face $\phi_+\lb\mathcal C''\rb$, and the contour of integration w.r.t. $z'$ is slightly deformed to the face $\phi_-\lb\mathcal C'\rb=\phi_-\lb\mathcal C''\rb$. Therefore the integration contour in
  \ben
  \sum_{\quad\mathcal C'\in\mathsf C_{\phi_-\lb \mathcal C''\rb}}\oint_{\mathcal C'} \frac{\bar{\Psi}_{\mathcal C',-}\lb z'\rb^{-1}dz'}{
  \lb z'-z\rb \lb z'' -z'\rb}
  \ebn 
  can be collapsed through the face $\phi_-\lb \mathcal C''\rb$ and the corresponding integral vanishes. The vanishing of the ``$-+$'' component may be shown in a similar way using in addition that, by definition of $\bar\Psi$, we have $\Psi_{\mathcal C,-}\bar\Psi_{\mathcal C,+}=
  \Psi_{\mathcal C,+}\bar\Psi_{\mathcal C,-}$ and shrinking the contours through positive faces.
  
  For the ``$++$'' component, the analog of the above is
   \begin{align*}
   &\Bigl( \mathcal P_{\Sigma,+}\sum\limits_{f\in\mathsf F_{+}}\sum\limits_{\mathcal C,\mathcal C'\in\mathsf C_f}
       \Psi_{\mathcal C,-}
          \Pi_{ \mathcal C\leftarrow\mathcal C',+}\Psi_{\mathcal C',-}^{\;-1} g_+\Bigr)\lb z\rb=\\&=
          \sum_{\mathcal C,\mathcal C''\in\mathsf C}\sum_{\;\;\mathcal C'\in\mathsf C_{\phi_+\lb \mathcal C''\rb}}\!\!\!\Pi_{\mathcal C,+}\frac{1}{\lb 2\pi i\rb^2}
     \oint_{\mathcal C''}     \oint_{\mathcal C'}
    \frac{
       \chi_{\mathcal C}\lb z\rb\Psi_{\mathcal C,+}\lb z\rb\bar\Psi_{\mathcal C,-}\lb z\rb
              \bar\Psi_{\mathcal C',+}\lb z'\rb^{\;-1}\Psi_{\mathcal C'',-}\lb z''\rb^{-1} 
        g_{\mathcal C'',+}\lb z''\rb dz'dz''}{\lb z'-z\rb \lb z'' -z'\rb}.      
   \end{align*}
  The contours of integration w.r.t. $z'$ and $z''$ are deformed to $\phi_-\lb \mathcal C'\rb$ and $\phi_-\lb \mathcal C''\rb$, and  the former is located to the positive side of the latter on the coinciding faces. Collapsing the contour in the integral
  \ben
  \sum_{\quad\mathcal C'\in\mathsf C_{\phi_+\lb \mathcal C''\rb}}\oint_{\mathcal C'}\frac{ \bar\Psi_{\mathcal C',+}\lb z'\rb^{\;-1} dz'}{\lb z'-z\rb\lb z''-z' \rb}
  \ebn
  through the face $\phi_+\lb \mathcal C''\rb$, we eventually pick up a residue at $z'=z$, equal to $\ds\frac{2\pi i \bar\Psi_{\mathcal C',+}\lb z\rb^{\;-1}}{z''-z}$, if $\mathcal C\in \mathsf C_{\phi_+\lb \mathcal C''\rb}$. Using this, reorganize the previous expression as
  \ben
   \sum_{\mathcal C''\in\mathsf C}\sum_{\;\;\mathcal C\in\mathsf C_{\phi_+\lb \mathcal C''\rb}}\Pi_{\mathcal C,+}\frac{1}{ 2\pi i}
       \oint_{\mathcal C''}     
      \frac{
         \chi_{\mathcal C}\lb z\rb\Psi_{\mathcal C,-}\lb z\rb\Psi_{\mathcal C'',-}\lb z''\rb^{-1} 
          g_{\mathcal C'',+}\lb z''\rb dz''}{ z'' -z}. 
  \ebn
  For $\mathcal C\ne \mathcal C''$, the integral defines a function of $z$ that analytically continues to $\varphi_+\lb \mathcal C''\rb\supset \phi_-\lb \mathcal C\rb$ and therefore vanishes under the action of $\Pi_{\mathcal C,+}$. There remains a sum
    \ben
     \sum_{\mathcal C\in\mathsf C}\Pi_{\mathcal C,+}\frac{1}{ 2\pi i}
         \oint_{\mathcal C}     
        \frac{
           \chi_{\mathcal C}\lb z\rb\Psi_{\mathcal C,-}\lb z\rb\Psi_{\mathcal C,-}\lb z''\rb^{-1} 
            g_{\mathcal C,+}\lb z''\rb dz''}{ z'' -z},
    \ebn
    where $\mathcal C$ is slightly deformed to $\phi_-\lb\mathcal C\rb$ as to avoid the singularity at $z''=z$. Deforming it instead to $\phi_+\lb\mathcal C\rb$ we obtain a function of $z$ annihilated by $\Pi_{\mathcal C,+}$ at the expense of picking up the residue at $z''=z$, equal to $g_{\mathcal C,+}$. This ultimately yields the expected result $\sum_{\mathcal C\in\mathsf C}\Pi_{\mathcal C,+}\chi_{\mathcal C}g_{\mathcal C,+}=g_+$. The calculation for the ``$--$'' component is  completely analogous.
  \epf
  
  \begin{theo}\label{theomultitau}
  Suppose that the functions $\Psi_{\mathcal C,\pm}\lb z\rb $ appearing in the direct factorization of individual jumps $J_{\mathcal C}\lb z\rb$ smoothly depend on an additional parameter $t$. If the solution $\bar\Psi_{\pm}\lb z\rb$ of the dual RHP exists and smoothly depends on $t$, then 
  \beq\label{widom_gen}
  \partial_t\ln\tau\left[J\right]=\sum_{\mathcal C\in\mathsf C}
  \frac1{2\pi i}\oint_{\mathcal C}\operatorname{Tr}
  \left\{J_{\mathcal C}^{-1}\partial_tJ_{\mathcal C}\left[\partial_z\bar\Psi_{\mathcal C,-}\, {\bar\Psi_{\mathcal C,-}}^{-1}+\Psi_{\mathcal C,+}^{-1}\,\partial_z\Psi_{\mathcal C,+}
    \right]\right\}dz.
  \eeq
  \end{theo}
  \pf We are going to mimick the proof of Theorem~\ref{WidomII}.
  Note e.g. that the operators $\mathsf A_{\pm\mp}$ in \eqref{taugen}, or more precisely their conjugates $\tilde{\mathsf A}_{\pm\mp}$ in \eqref{taugen2}, are analogs of the operators
    $\Pi_+J^{-1}\Pi_-$ and $\Pi_- J\Pi_+$.
   The main difference here is that it becomes convenient to write various projection operators as explicit contour integrals. 
   
  Differentiating the determinant yields
  \begin{align*}
  \partial_t\ln\tau\left[J\right]=\operatorname{Tr}_H\Bigl(\lb\mathbb 1+L\rb^{-1}\partial_t L\Bigr)=
  \operatorname{Tr}_H\lb \mathcal P_{\Sigma,-}\bigl|_{H_+}\partial_t\mathsf A_{+-}  +  \mathcal P_{\Sigma,+}\bigl|_{H_-} \partial_t\mathsf A_{-+} \rb=\operatorname{Tr}_H\lb
  \mathcal P_{\Sigma,-}\partial_t\mathsf A_{+-} +  \mathcal P_{\Sigma,+} \partial_t\mathsf A_{-+} \rb,
  \end{align*}
  where the last equality is obtained using that  $\operatorname{im}\mathsf A_{\pm \mp}\subseteq H_{\pm}$. Moreover, thanks to the property that $ H_{\pm}\subseteq\operatorname{ker}\mathsf A_{\pm \mp}$, the first projector in the definition (\ref{psigpm}) of $\mathcal P_{\Sigma,\pm}$ may be omitted. The computation of traces then reduces to residue calculation. Indeed, we have
  \begin{align*}
  \begin{aligned}
  \operatorname{Tr}_H \mathcal P_{\Sigma,-}\partial_t\mathsf A_{+-}
  =\!\!\!\sum_{\mathcal C,\mathcal C''\in\mathsf C}\sum_{\;\;\mathcal C'\in\mathsf C_{\phi_{-}\lb \mathcal C''\rb}}
  \frac{1}{\lb 2\pi i\rb^2}\oint_{\mathcal C'}\oint_{\mathcal C''}
  \operatorname{Tr}\Biggl\{&\,
  \frac{\chi_{\mathcal C}\lb z\rb\Psi_{\mathcal C,+}\lb z\rb\bar\Psi_{\mathcal C,-}\lb z\rb
            \bar\Psi_{\mathcal C',-}\lb z'\rb^{\;-1}\Psi_{\mathcal C',+}\lb z'\rb^{-1}}{z'-z}\times\\
            \times&\,\frac{\partial_t\lb
  \Psi_{\mathcal C',+}\lb z'\rb \Psi_{\mathcal C'',+}\lb z''\rb^{-1}\rb}{z''-z'}\Bigl|_{z''=z}\Biggr\}\; dz'dz=
  \end{aligned}\\
  =\sum_{\mathcal C\in\mathsf C}\sum_{\;\;\mathcal C'\in\mathsf C_{\phi_{-}\lb \mathcal C\rb}}
    \frac{1}{\lb 2\pi i\rb^2}\oint_{\mathcal C}\oint_{\mathcal C'}
    \frac{\operatorname{Tr}\bar\Psi_{\mathcal C,-}\lb z\rb
                \bar\Psi_{\mathcal C',-}\lb z'\rb^{\;-1}\left[
                \Psi_{\mathcal C,+}\lb z\rb^{-1}\partial_t\Psi_{\mathcal C,+}\lb z\rb- \Psi_{\mathcal C',+}\lb z'\rb^{-1}\partial_t \Psi_{\mathcal C',+}\lb z'\rb\right] dz dz'}{\lb z'-z\rb^2}.
  \end{align*}
  Recall that the contours $\mathcal C'$ of integration with respect to $z'$ are slightly pushed to positive faces $\phi_+\lb \mathcal C'\rb$ according to the definition of $\mathcal P_{\Sigma,-}$. The contribution of the 1st term under trace is readily computed by collapsing the contours $\mathcal C'$ through negative faces and is given by (minus) the residue at $z'=z$,
  \begin{subequations}
  \beq\label{sum01}
  \frac{1}{2\pi i}\sum_{\mathcal C\in\mathsf C}\oint_{\mathcal C}
  \operatorname{Tr}\left\{
  \partial_z\bar\Psi_{\mathcal C,-} \bar\Psi_{\mathcal C,-}^{-1}\Psi_{\mathcal C,+}^{-1}\partial_t \Psi_{\mathcal C,+}\right\}\,dz.
  \eeq
  To compute in a similar way the 2nd term under trace, rearrange the sum $\sum\limits_{\mathcal C\in\mathsf C}\sum\limits_{\;\;\mathcal C'\in\mathsf C_{\phi_{-}\lb \mathcal C\rb}}$ as $\sum\limits_{\mathcal C'\in\mathsf C}\sum\limits_{\;\;\mathcal C\in\mathsf C_{\phi_{-}\lb \mathcal C'\rb}}$. Shrinking afterwards the contours $\mathcal C$ through negative faces we meet no poles and therefore the corresponding integrals sum up to zero.
  
  One can analogously prove that
  \beq\label{sum02}
   \operatorname{Tr}_H \mathcal P_{\Sigma,+}\partial_t\mathsf A_{-+}
    =-\frac{1}{2\pi i}\sum_{\mathcal C\in\mathsf C}\oint_{\mathcal C}
      \operatorname{Tr}\left\{
      \partial_z\bar\Psi_{\mathcal C,+} \bar\Psi_{\mathcal C,+}^{-1}\Psi_{\mathcal C,-}^{-1}\partial_t \Psi_{\mathcal C,-}\right\}\,dz.
  \eeq
  \end{subequations}
  It remains to show that the sum of (\ref{sum01}) and (\ref{sum02}) coincides with the right side of \eqref{widom_gen}. To this end,  note that
  \begin{align*}
  &\operatorname{Tr}J_{\mathcal C}^{-1}\partial_tJ_{\mathcal C}\left[\partial_z\bar\Psi_{\mathcal C,-}\, {\bar\Psi_{\mathcal C,-}}^{-1}+\Psi_{\mathcal C,+}^{-1}\,\partial_z\Psi_{\mathcal C,+}
      \right]=\\
      =&\,\operatorname{Tr}\left\{\lb\partial_t\Psi_{\mathcal C,+}-\partial_t\Psi_{\mathcal C,-}\, \Psi_{\mathcal C,-}^{-1}\Psi_{\mathcal C,+}\rb\partial_z\bar\Psi_{\mathcal C,-} \bar\Psi_{\mathcal C,-}^{-1}\Psi_{\mathcal C,+}^{-1}+\lb \partial_t\Psi_{\mathcal C,+}\,\Psi_{\mathcal C,+}^{-1}-\partial_t\Psi_{\mathcal C,-}\,\Psi_{\mathcal C,-}^{-1}\rb\partial_z\Psi_{\mathcal C,+}\,\Psi_{\mathcal C,+}^{-1}\right\}=\\
      =&\,\operatorname{Tr}\left\{\Psi_{\mathcal C,+}^{-1}\partial_t\Psi_{\mathcal C,+}\lb \partial_z\bar\Psi_{\mathcal C,-} \bar\Psi_{\mathcal C,-}^{-1}+\Psi_{\mathcal C,+}^{-1}\partial_z\Psi_{\mathcal C,+}\rb-\Psi_{\mathcal C,-}^{-1}\partial_t\Psi_{\mathcal C,-}\lb\Psi_{\mathcal C,-}^{-1}\partial_z\Psi_{\mathcal C,-}+\partial_z\bar{\Psi}_{\mathcal C,+}\bar{\Psi}_{\mathcal C,+}^{-1}\rb\right\},
  \end{align*}
  where to obtain the first equality, we use that $J_{\mathcal C}=\Psi_{\mathcal C,-}^{-1}\Psi_{\mathcal C,+}$; the second equality is obtained by replacing $\Psi_{\mathcal C,+}=\Psi_{\mathcal C,-}\bar{\Psi}_{\mathcal C,+}\bar{\Psi}_{\mathcal C,-}^{-1}$ in the 4th term under trace. In the last expression, the 1st and 4th term reproduce (\ref{sum01}) and (\ref{sum02}). The 2nd and 3rd term are given by the boundary values of functions analytic in $\varphi_+\lb\mathcal C\rb$ and  $\varphi_-\lb\mathcal C\rb$, therefore the corresponding integrals vanish.
   \epf
   
  \begin{rmk} 
  We conclude this subsection by mentioning the work of Palmer \cite{Palmer} on the tau function of the massive Euclidean Dirac operator in the presence of Aharonov-Bohm fluxes. While it may seem unrelated to the present paper, it is the adaptation of the localization ideas of  \cite{Palmer} to the chiral case which led to \cite{GL16}. Here we made a substantial further improvement by getting rid of artificial doubling of the RHP contours, generalizing the results to arbitrary oval configurations and providing a concise definition of $\tau\left[J\right]$. It might be interesting to understand whether it is possible to go backwards and apply our results, in particular, series expansions of Subsection~\ref{subseccombi}, to the study of correlation functions of twist fields in free massive QFTs. 
  \end{rmk}

  \subsection{Jimbo-Miwa-Ueno differential}
  In this subsection, we explain how to recover from Theorem~\ref{theomultitau} the Jimbo-Miwa-Ueno definition \cite{JMU} of the  isomonodromic tau function for systems of linear differential equations with rational coefficients.
  Let us start with a Fuchsian system with $n$ regular singular points $a_0=0,a_1,\ldots,a_{n-2},a_{n-1}=\infty$ on $\Cb\Pb^1$, 
  \beq\label{FS}
  \partial_z\Phi=\Phi A\lb z\rb,\qquad A\lb z\rb = \sum_{k=0}^{n-2}\frac{A_{k}}{z-a_{k}},\qquad 
  A_{k}\in\operatorname{Mat}_{N\times N}\lb \Cb\rb.
  \eeq
  For simplicity it is assumed that $a_1,\ldots,a_{n-2}\in \mathbb R_{>0}$ and that the singularities are ordered as $a_1<\ldots<a_{n-2}$. The fundamental solution $\Phi\lb z\rb$ can then be considered as a single-valued analytic function on $\Cb\backslash\mathbb{R}_{\ge0}$. Similarly to Subsection~\ref{subsec4reg}, we also assume that $A_0,\ldots,A_{n-2},A_{n-1}:=-\sum_{k=0}^{n-2}A_{k}$ are diagonalizable as $A_{k}=G_{k}^{-1}\Theta_{k} G_{k}$ and have non-resonant eigenvalues, so that in the neigborhood of each singular point we have
  \ben
  \Phi\lb z\rb= C_{k,\epsilon} \lb a_{k}-z\rb^{\Theta_{k}}G^{\lb k\rb}\lb z\rb, \qquad \epsilon =\operatorname{sgn} \Im z,
  \ebn
  where $G^{\lb k\rb}\lb z\rb$ are holomorphic invertible and normalized so that $G^{\lb k\rb}\lb a_{k}\rb=G_{k}$ (for $a_{n-1}=\infty$ the formula above should be modified in the obvious manner). The connection matrices $\left\{C_{k,\pm}\right\}$ satisfy the compatibility conditions analogous to \eqref{compa} and, together with local monodromy exponents $\left\{\Theta_{k}\right\}$, encode the monodromy representation of $\pi_1\lb\Cb\Pb^1\backslash\left\{n\text{ points}\right\}\rb$ associated to $\Phi\lb z\rb$.

\begin{figure}[h!]
     \centering
     \includegraphics[height=6cm]{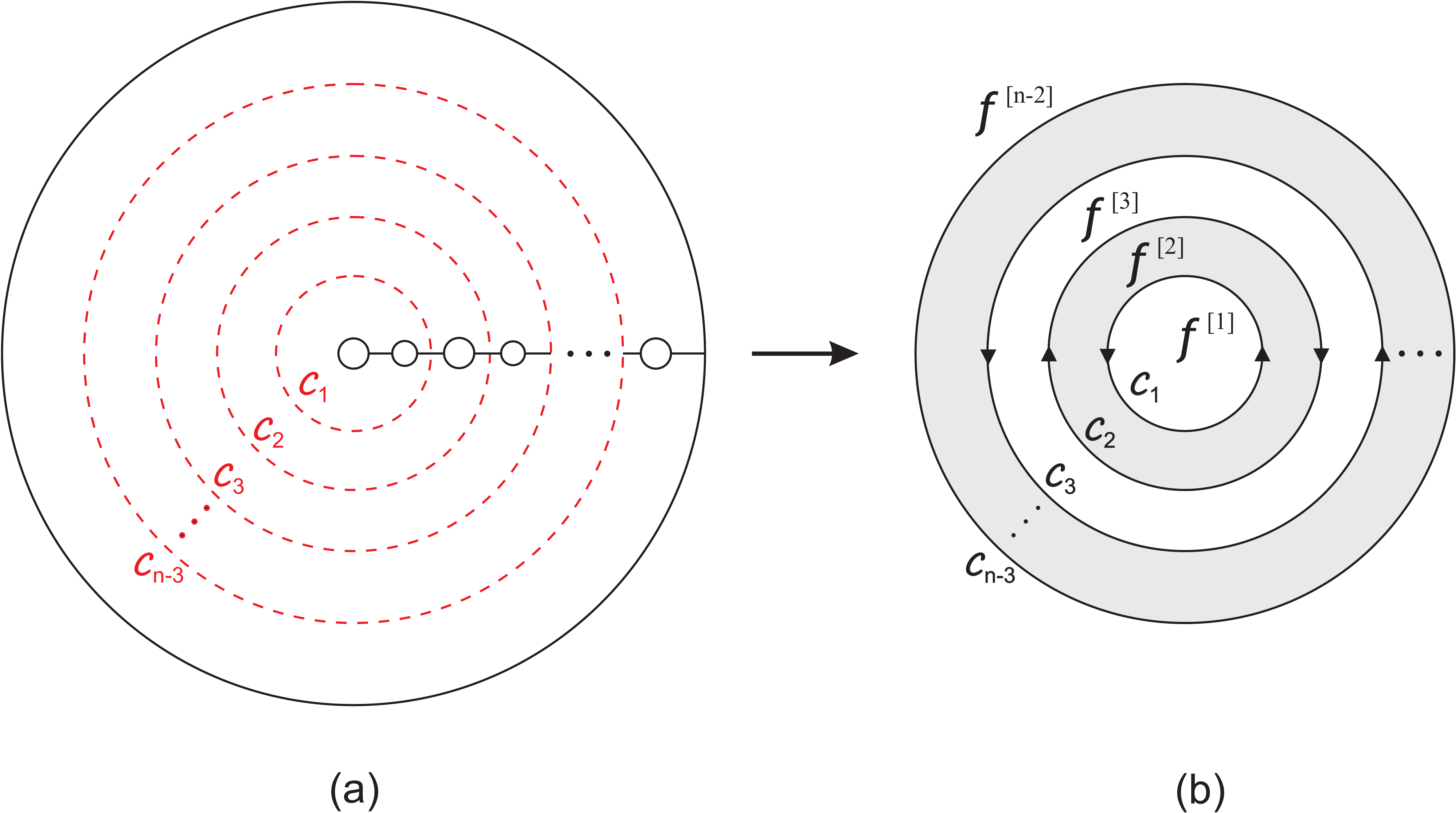}
     \begin{minipage}{0.6\textwidth}
     \caption{RH contour and coloring by $\left\{+,-\right\}$ associated to linear pants decomposition for Fuchsian systems.
                        \label{fig_gar}}
                                         \end{minipage}
                       \end{figure}

  Different pants decompositions of the $n$-punctured sphere give rise to different RHPs associated with the linear system \eqref{FS} and distinct Fredholm determinant representations of the tau functions, adapted to analysis of different asymptotic regimes. Since at this point we only want to give an example of an $n>4$ analog of the relation \eqref{tau4tauJMU}, let us pick the simplest ``linear''\footnote{One may assign to an arbitrary collection of non-intersecting ovals a tree graph with vertices given by faces in $\mathsf F_{+}\cup \mathsf F_{+}$ and the edges given by their common boundaries. The contour shown in Fig.~\ref{fig_gar}b leads to a linear graph whereas e.g. the contour in Fig.~\ref{fig_sic}c yields a star-shaped graph with 4 vertices.}
  pants decomposition leading to a RHP set on a collection of $n-3$ circles $\mathcal C_1,\ldots,\mathcal C_{n-3}$ decomposing $\Cb\Pb^1$ into $n-2$ faces $f^{[1]},\ldots,f^{[n-2]}$ as shown in Fig.~\ref{fig_gar}. By convention, the faces $f^{[2k+1]}$ and $f^{[2k]}$ will be of color $+$ and $-$, respectively. 
  
    Let $M_{k}\in\mathrm{GL}\lb N,\Cb\rb$ ($k=0,\ldots,n-1$) denote the monodromy of $\Phi\lb z\rb$ along the contour starting on the negative real axis and going around $a_{k}$ counterclockwise. These monodromies satisfy the cyclic relation $M_0M_1\ldots M_{n-1}=\mathbb 1$. It will be convenient for us to consider the products $M_{0\to k}:=M_0\ldots M_k$ and suppose that they can be diagonalized as 
    \ben
    M_{0\to k}=S_k^{-1} e^{2\pi i\mathfrak S_k}S_k, \qquad k=0,\ldots, n-2,
    \ebn
    where the eigenvalues of diagonal matrices $\mathfrak S_k$ are assumed to be pairwise distinct $\operatorname{mod}\;\Zb$. It may also be assumed that $\operatorname{Tr}\mathfrak S_k=\sum_{j=0}^k \operatorname{Tr}\Theta_{j}$ and that $\mathfrak S_0=\Theta_0$, $\mathfrak S_{n-2}=-\Theta_{n-1}$.
  
  Denote by $\Phi^{[k]}\lb z\rb$ ($k=1,\ldots,n-2$) the solution of 3-point Fuchsian system associated to the face $f^{[k]}$ (cf Subsection \ref{subsec4reg}) which has regular singularities at $0$, $a_k$ and $\infty$ characterized by monodromies $M_{0\to k-1}$, $M_{k}$ and 
  $M_{0\to k}^{-1}$. The local behavior of this solution near the singular points is given by
  \ben
  \Phi^{[k]}\lb z\rb =
  \begin{cases}
  S_{k-1} \lb -z\rb^{\mathfrak S_{k-1}} G_0^{[k]}\lb z\rb,\qquad & z\to0, \\
  C_{k,\epsilon}\lb a_k-z\rb^{\Theta_k}G_{k}^{[k]}\lb z\rb,\qquad
  & z\to a_k,\quad\epsilon=\mathrm{sgn}\Im z, \\
  S_{k} \lb -z\rb^{\mathfrak S_{k}} G_{\infty}^{[k]}\lb z\rb,\qquad & z\to \infty,
  \end{cases}
  \ebn
  where $G^{[k]}_{0,k,\infty}\lb z\rb$ are holomorphic invertible in the respective neighborhoods of $0,a_k,\infty$. The 3-point solutions define the jumps on $\mathcal C_1\cup\ldots\cup\mathcal C_{n-3}$: in the notation of the previous subsection, we have
 \beq\label{local}
  \begin{aligned}
  &\Psi_{\mathcal C_{2k-1},+}\lb z\rb=G_0^{[2k]}\lb z\rb = \lb-z\rb^{-\mathfrak S_{2k-1}}S_{2k-1}^{-1}\Phi^{[2k]}\lb z\rb,\\ 
  &\Psi_{\mathcal C_{2k-1},-}\lb z\rb = G_{\infty}^{[2k-1]}\lb z\rb = \lb-z\rb^{-\mathfrak S_{2k-1}}S_{2k-1}^{-1}\Phi^{[2k-1]}\lb z\rb,\\
  &\Psi_{\mathcal C_{2k},+}\lb z\rb = G_{\infty}^{[2k]}\lb z\rb = \lb-z\rb^{-\mathfrak S_{2k}}S_{2k}^{-1}\Phi^{[2k]}\lb z\rb,\\
  &\Psi_{\mathcal C_{2k},-}\lb z\rb = G_{0}^{[2k+1]}\lb z\rb = \lb-z\rb^{-\mathfrak S_{2k}}S_{2k}^{-1}\Phi^{[2k+1]}\lb z\rb,
  \end{aligned}
  \eeq
  and $\bar{\Psi}\lb z\rb = \Phi^{[k]}\lb z\rb^{-1}\Phi\lb z\rb$ for $z\in f^{[k]}$. 
  
  Substitute these expressions into differentiation formula (\ref{widom_gen}) choosing therein  $t=a_k$. The only circles that contribute to the sum are $\mathcal C_{k-1}$ and $\mathcal C_k$ (for the others $\partial_t J=0$). Moreover, for instance, for odd $k$ the only $\Psi_{\mathcal C,\epsilon}$ depending on $a_k$ are $\Psi_{\mathcal C_k,-}$ and $\Psi_{\mathcal C_{k-1},-}$, so that in this case
  \begin{align*}
   &\partial_{a_{k}}\ln\tau\left[J\right]=\\
   =&\,-\frac1{2\pi i}\left[\oint_{\mathcal C_{k-1}}\!\!\!\!\!\! \operatorname{Tr}\left\{
         \partial_z\bar\Psi_{\mathcal C_{k-1},+} \bar\Psi_{\mathcal C_{k-1},+}^{-1}\Psi_{\mathcal C_{k-1},-}^{-1}\partial_{a_k} \Psi_{\mathcal C_{k-1},-}\right\}dz+\oint_{\mathcal C_{k}} \!\!\!\!\operatorname{Tr}\left\{
                  \partial_z\bar\Psi_{\mathcal C_{k},+} \bar\Psi_{\mathcal C_{k},+}^{-1}\Psi_{\mathcal C_{k},-}^{-1}\partial_{a_k} \Psi_{\mathcal C_{k},-}\right\}dz\right]=\\
                  =&\,-\operatorname{res}_{z=a_k}\operatorname{Tr}\left\{\partial_z\lb{\Phi^{[k]}}^{-1}\Phi\rb
                  \Phi^{-1}\partial_{a_k}\Phi^{[k]}\right\}=\\
                  =&\,\operatorname{res}_{z=a_k}\operatorname{Tr}\left\{\partial_z\lb{G^{[k]}_k}^{-1}G^{(k)}\rb
                                    {G^{(k)}}^{-1}\lb\frac{\Theta_k}{z-a_k}G^{[k]}_k-\partial_{a_k}G^{[k]}_k\rb\right\}=\\
  =&\,  \operatorname{res}_{z=a_k}\operatorname{Tr}\left\{\lb\partial_z G^{(k)}\,{G^{(k)}}^{-1}-\partial_z G_k^{[k]}\,{G_k^{[k]}}^{-1}\rb\frac{\Theta_k}{z-a_k}\right\}      =\\
  =&\, \partial_{a_k}\tau_{\mathrm{JMU}}-\partial_{a_k}\tau_{\mathrm{JMU}}^{[k]},                          
  \end{align*}
  where $\tau_{\mathrm{JMU}}$ is the tau function of the $n$-point  system \eqref{FS} and $\tau_{\mathrm{JMU}}^{[k]}=\mathrm{const}\cdot a_k^{\frac12\operatorname{Tr}\lb\mathfrak S_k^2-\mathfrak S_{k-1}^2-\Theta_k^2\rb}$ is the tau function of the $3$-point Fuchsian system for $\Phi^{[k]}$. The transition from the 2nd to the 3rd line is obtained using that the integrands continue to the same meromorphic function on $f^{[k]}$ with the only pole at $z=a_k$. The 4th line follows from \eqref{local} and the 5th from the fact that ${G^{[k]}_k}^{-1}G^{(k)}$,  ${G^{(k)}}^{-1}$ and $\partial_{a_k}G^{[k]}_k$ are holomorphic in $f^{[k]}$. The final equality follows from the Jimbo-Miwa-Ueno definition of the tau function, cf \cite[eq. (1.23)]{JMU}. 
  
  We have thus shown that for Fuchsian systems and linear pants decomposition the tau function defined by the Fredholm determinant \eqref{taugen} coincides with
  \beq
  \tau\left[J\right]=\frac{\tau_{\mathrm{JMU}}\lb a_1,\ldots,a_{n-2}\rb}{\prod_{k=1}^{n-2}
  \tau_{\mathrm{JMU}}^{[k]}\lb a_k\rb}.
  \eeq
  When the singular points $a_0,\ldots a_{n-1}$ are irregular, one can obtain a similar identification by decomposing the original RHP into e.g. an $n$-point Fuchsian one (with regular singularities at $a_0,\ldots,a_{n-1}$) and $n$ two-point RHPs with one regular and one irregular singularity.

     \vspace{0.3cm}
      
        \noindent
        { \small \textbf{Acknowledgements}.  The authors would like to thank M. Bertola, T. Grava, Y. Haraoka, N. Iorgov, A. Its, K. Iwaki, H. Nagoya, A. Prokhorov and V. Roubtsov for useful discussions. The present work was supported by the PHC Sakura grant No. 36175WA and CNRS/PICS project ``Isomonodromic deformations and conformal field theory''. The work of P.G. was partially supported the Russian Academic Excellence Project `5-100' and by the RSF grant No. 16-11-10160. In particular, results of Subsection 3.1 have been obtained using support of Russian Science Foundation. P.G. is a Young Russian Mathematics award winner and would like to thank its sponsors and jury. }

 \end{document}